\documentclass[10pt, a4paper, onecolumn]{article}

\usepackage{arxiv}

\usepackage[utf8]{inputenc} 
\usepackage[T1]{fontenc}    
\usepackage[breaklinks]{hyperref}
\usepackage{url}
\usepackage{booktabs}       
\usepackage{amsfonts}       
\usepackage{nicefrac}       
\usepackage{microtype}      
\usepackage{amsmath}
\usepackage{amsfonts}
\usepackage{cite}
\usepackage{amssymb}
\usepackage{amsthm}
\usepackage{verbatim}
\usepackage{enumerate}
\usepackage[normalem]{ulem}
\usepackage{mdwlist} 
\usepackage{color}
\usepackage[dvipsnames]{xcolor}
\usepackage[utf8]{inputenc}

\usepackage{algpseudocode}
\usepackage{algorithmicx}
\usepackage{lineno}
\usepackage{framed,multirow}
\usepackage[final]{pdfpages}
\usepackage{latexsym}
\usepackage{wrapfig}

\usepackage{graphicx, subfig}
\usepackage{verbatim}
\usepackage{bm}
\usepackage{authblk}
\expandafter\def\expandafter\UrlBreaks\expandafter{\UrlBreaks
    \do\a\do\b\do\c\do\d\do\e\do\f\do\g\do\h\do\i\do\j
    \do\k\do\l\do\m\do\n\do\o\do\p\do\q\do\r\do\s\do\t
    \do\u\do\v\do\w\do\x\do\y\do\z\do\A\do\B\do\C\do\D
    \do\E\do\F\do\G\do\H\do\I\do\J\do\K\do\L\do\M\do\N
    \do\O\do\P\do\Q\do\R\do\S\do\T\do\U\do\V\do\W\do\X
    \do\Y\do\Z\do\/\do-}

\title{Random Matrix Analysis of Multiplex Networks}

\author[1]{Tanu Raghav}
\author[1]{Sarika Jalan}

\affil[1]{Complex Systems Lab, Discipline of Physics, Indian Institute of Technology Indore, Khandwa Road, Simrol, Indore 453552, India}

\begin{document}
\maketitle

\begin{abstract}
We investigate the spectra of adjacency matrices of multiplex networks under random matrix theory (RMT) framework. Through extensive numerical experiments, we demonstrate that upon multiplexing two random networks, the spectra of the combined multiplex network exhibit superposition of two Gaussian orthogonal ensemble (GOE)s for very small multiplexing strength followed by a smooth transition to the GOE statistics with an increase in  the multiplexing strength. Interestingly, {\it randomness} in the connection architecture, introduced by random rewiring to 1D lattice, of at least one layer may govern nearest neighbor spacing distribution (NNSD) of the entire multiplex network, and in fact, can drive to a transition from the Poisson to the GOE statistics or vice versa. Notably, this transition transpires for a very small number of the random rewiring corresponding to the small-world transition. Ergo, only one layer being represented by the small-world network is enough to yield GOE statistics for the entire multiplex network. 
Spectra of adjacency matrices of underlying interaction networks have been contemplated to be related with dynamical behaviour of the corresponding complex systems, the investigations presented here have implications in achieving better structural and dynamical control to the systems represented by multiplex networks against structural perturbation in only one of the layers.

\end{abstract}




\section{Introduction}
To anticipate and understand various properties of real-world complex systems, networks furnish a simple framework by considering a complex system in terms of nodes and interactions. In various real-world complex systems such as biological (e.g., food-web \cite{food}, nervous system \cite{nervous}, cellular metabolism \cite{cel_meta}, protein-protein interaction network \cite{ppi}, gene regulatory networks \cite{gene}), social (e.g., scientific collaboration \cite{collab}, citation \cite{cite}), linguistic \cite{linguistics}, and technological (e.g., Internet \cite{net}, power-grid \cite{power_grid} etc), properties of underlying network structures have been investigated to predict their dynamical behaviours. Networks framework relies on modeling such real-world complex systems by representing them in terms of nodes (interacting units) and links between these nodes. Such assemblage can also account for intricate structural and/or dynamical information of the system arising due to different types of relationships shared by the same units. For example, in a transport system, cities can form different layers of a multilayer network depending on their connectivity in terms of the mode of travel such as rail, bus, and airlines \cite{air, trans}. Another example of multilayer network is social systems where people can connect through different online social platforms such as Facebook, Twitter, etc. \cite{ML} forming different layers of the social multilayer networks. Another example having multiple layers of connections is that of a brain where interacting neurons can either be realized in a form of a physical network or can correspond to a network which captures a functional relationship between them \cite{brain}. 
To have a better and more precise understanding of such systems having different types of interactions among the same units, the concept of multilayer networks \cite{ML1, ML2, ML3, ML4, ML5} was coined. One of the simplest forms of this framework is a multiplex network where each node is simultaneously present in all its layers. The inter-layer connections between the layers, referred to as multiplexing strength, provide a quantitative measure of the impact of dynamical and/or structural properties of one layer on another layer and vice versa.

\begin{figure}
\includegraphics[width=1\textwidth]{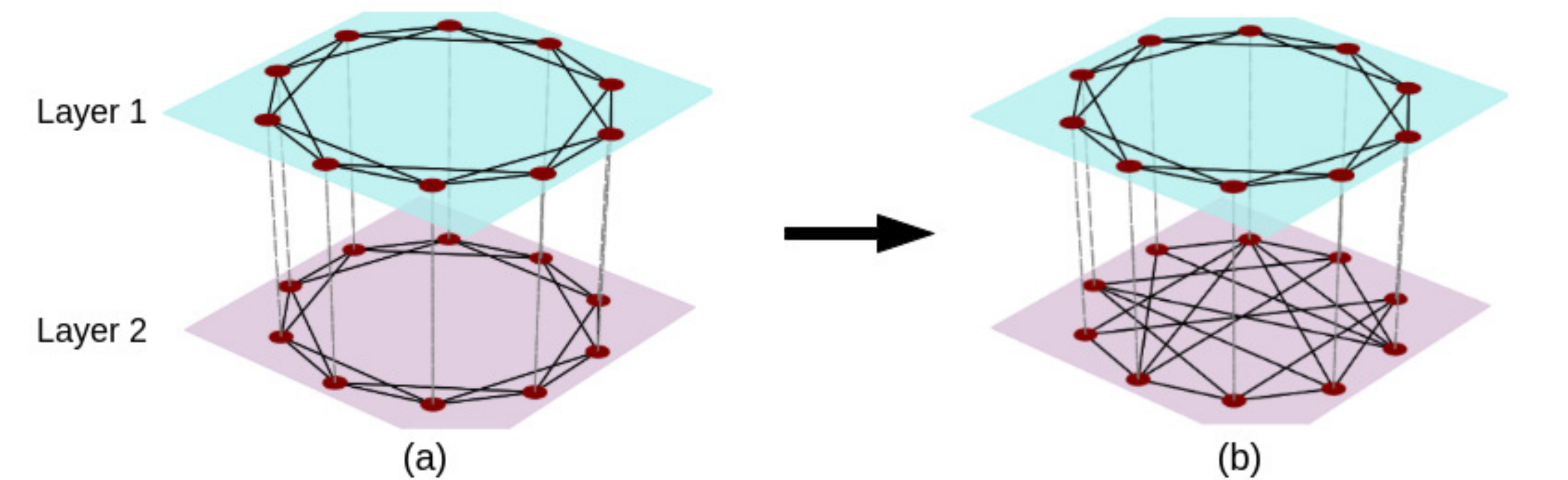}
\caption{ Schematic diagram representing a bi-layer multiplex network. (a) Both layers are represented by 1D lattice with $k$ nearest neighbour connections. (b) Edges are rewired in the second layer with a probability $p_r$. }
\label{fig1}
\end{figure}

Furthermore, voluminous literature on multilayer networks demonstrates that the optimization in one layer can affect the behavior of the whole multilayer network. For example, rewiring in only single layer can attain localization of the multilayer network consisting of that layer \cite{prio, loc1}. An optimization of synchronizability has been shown to be achieved in a multilayer network by rewiring only one of its layers \cite{sanjiv, sync1}. These examples demonstrate the impact of structural properties of one layer on emergent behaviors of the entire multilayer systems as well as of the individual layers. 
Therefore, it is important to understand the impact of structural changes as a consequence of rewiring in one of the layers on various properties of the multilayer network at a more fundamental level. Also, various real-world complex systems that comprise millions of agents interacting with each other have properties that are neither completely random nor regular, instead, they possess nontrivial characteristics that indicate a more elaborate or more complex organization \cite{power_grid, social}.

This paper studies the spectra of multiplex networks consisting of two layers and investigates the effect of (i) rewiring the connections in one layer and (ii) change in the multiplexing strength on various spectral properties of the entire multiplex network under the RMT framework. Besides achieving fundamental understanding of the impact of structural randomness in one layer on the spectra of entire multiplex networks, this kind of setup has practical implications as well. There exist diverse factors that can account for disrupting the connection architecture of one or more layers of a system represented by the multilayer network. For example, in a transport system, catastrophe or an accident can affect one mode of transport (e.g., railway or airways mode) making the connections route to the corresponding part of the network in one layer more random, which may lead to affect response of other modes eventually affecting the response of the entire transport system \cite{random, congestion}. In the example of social networks, fake profiles and account hacking are one of the biggest threats to social platforms, which can affect the architecture of the entire social network \cite{social1, social2, social3}. 

Also, it has been shown that how multiplexing strength, i.e., how properties of one layer can be affected by dynamical and/or structural properties of another layer and vice versa, which in turn can affect the dynamical response to the entire system. An extensive literature on multilayer networks has demonstrated the importance of inter-layer connection strengths in deciding a response of the underlying system or individual layer \cite{dx1, dx2, dx3}. How strongly layers are multiplexed plays a crucial role in determining the structural and dynamical importance of nodes in their individual layer. For example, a hub node in one layer may be isolated or may have very less number of connections in the other layers or vice versa, and a node's importance or behavior in one layer may strongly get affected by the structural properties and importance of the same node in the other layer \cite{ML6}. Also, in the disease spreading models, the threshold of disease outbreak may be very different in the multilayer framework as compared to that of the single-layer network's framework \cite{ML1}. As already described, in the transport multilayer networks with each layer representing the mode of connectivity or a transportation mode, the strength of inter-layer coupling or multiplexing strength represents the impact of one mode of interactions on the other. For example, the city having an airport far located or having less frequent flight connections than those of the railway station or a bus stop, it is difficult to switch to the airline mode from the other modes. Therefore, a disruption in the flight connectivity involving such cities will have less impact on the other modes, representing low multiplexing (albeit locally involving mirror nodes corresponding to such cities). Additionally, there exist various social platforms which cost for creating an account, in this case also switching from one social platform (free of cost) to another (not free) is difficult, leading to less impact of disturbance in one platform on others yielding low multiplexing whereas switching from free to other free platform is easy which is the example of strong multiplexing. The multiplexing strength have been shown to have crucial impact of dynamical behavior of nodes in different layers \cite{dx1, dx2, Multi, Multi1}. Therefore, the second part of the paper revolves around impact of the multiplexing strength in combination with various network architecture of individual layers on various spectral properties of the entire multiplex network.

Further, an extensive amount of work has been carried out on understanding the spectra of graphs \cite{graph} relating various structural properties with the underlying network structure \cite{str_prop1, str_prop2, dismantle}, as well as dynamical processes on those networks \cite{dyn_prop2, nine}. For example, in the networks of coupled phase oscillators, a transition from an incoherence to a coherence phase occurs at the coupling strength determined by the largest eigenvalue of the underlying adjacency matrices \cite{dyn_prop1}. Also, in the case of virus spread, the epidemic threshold for virus propagation is equal to the inverse of the largest eigenvalue \cite{virus}. Furthermore, the eigenvalues of the adjacency matrix can be used to count the number of walks of a given length \cite{spectra2}. Graph spectra consists of signatures of various topological properties of underlying networks. For instance, if the eigenvalues of adjacency matrix are symmetric with respect to 0, the corresponding graph is bipartite, whereas, for the complete graph $K_N$ on $N$ nodes the largest eigenvalue $\lambda_1 = (N-1)$ and $\lambda_{2} =\lambda_{3} =........\lambda_{N} = -1$ \cite{spectra1}. Furthermore, degeneracy in eigenvalues correspond to partial and complete duplicates of the nodes \cite{Dup_nodes1, Dup_nodes2}.
The spectrum of a network consisting of $N$ nodes is the set of eigenvalues of its adjacency matrix $A$ and is denoted as [$\lambda$]. The spectral density then can be written as
\begin{equation}\label{eq1}
\rho(\lambda)=\dfrac{1}{N}\sum_{i=1}^{N}\delta(\lambda-\lambda_{i})
\end{equation} 
where $\delta(x)$ is Dirac's delta function.
Investigations on the density distribution of eigenvalues of the adjacency matrix of various networks have demonstrated that spectral density of the adjacency matrix of Erd\H{o}s-R\'{e}nyi (ER) random networks follow the semicircular law \cite{str_prop1, spectra2, sj1}. Whereas the spectral densities of various other network models, as well as real-world networks, are not semicircular; instead, they have some specific features depending on the minute details of the networks \cite{spectral_density}. For example, the small-world (SW) model networks show complex structure for the spectral densities with many sharp peaks, while spectral densities of scale-free model networks exhibit triangular distribution \cite{str_prop1,sj1}. Also, much attention has been focused on eigenvector localization due to its potential to provide understanding to behaviour of corresponding complex systems. For example, eigenvector localization properties have been related to the scaling parameter of scale-free networks \cite{Ev1}, as well as has been used for detecting communities in multilayer and temporal networks \cite{Ev2}. Further, eigenvector localization as well as it implications in multiplex networks due to rewiring of edges (in one of the layers) has been investigated \cite{prio}. Eigenvector localization of multiplex networks has been also studied for different weights of the inter-layer coupling \cite{dx3, Ev4}.

Here, we investigate the spectra of multiplex networks within the framework of random matrix theory (RMT). RMT, originally proposed to explain the statistical properties of nuclear spectra, was successful in predicting the properties of many different complex systems coming from a diverse range of fields such as disordered, quantum chaotic, and large complex atoms \cite{system}. The spectral density of random matrices, whose elements are Gaussian distributed random numbers, follows the Wigner’s semicircular law \cite{Mehta}. Spectra of various model networks and also real-world networks have been analyzed under RMT framework \cite{sj1, sj2}. 
A common practice in RMT is to study eigenvalue fluctuations via nearest neighbor spacing distribution (NNSD). It obeys two universal properties depending upon the underlying correlations among the eigenvalues. For correlated eigenvalues, NNSD follows the Wigner-Dyson formula of Gaussian orthogonal ensemble (GOE) statistics of RMT, this property is shown by real symmetric random matrices. On the other hand, for uncorrelated eigenvalues, NNSD follows Poisson statistics of RMT, which is a property followed by random matrices having Gaussian distributed random elements only along its diagonals.
Earlier RMT analysis of single layer networks show that despite having differences in model networks (in terms of various local and global properties), fluctuations of the eigenvalues of the adjacency matrices follow the universal predictions of the GOE ensemble of RMT. Furthermore, it has been shown that the NNSD of random networks, scale-free networks, and SW networks follow GOE statistics. Also, during the transition from the 1D lattice to the SW networks, it was found that for very small values of rewiring probability \textbf{($p$)}, rendering the initial 1D structure almost intact, NNSD follows Poisson statistics, whereas for $p=1$ it follows the GOE statistics. For $0\le p \le1$, NNSD shows an intermediate statistics between the Poisson and the GOE \cite{sj1, sj2}. However, all the investigations on various spectral properties of adjacency matrices are confined to single layer networks only \cite{RMT_geometric, RMT_powergrid, RMT_RGG, RMT_Normal_mode, RMT_brain, RMT_skeleton, RMT_quantum, RMT_random, RMT_ADHD, RMT_ginibre, RMT_limb, RMT_functional}.

The current paper focuses on the following: (i) how {\it randomness} in one layer can affect or govern the spectral properties of the entire multiplex networks. To understand this, we first consider two extreme cases of the network topologies; the 1D ring lattices with $k$ nearest neighbors interactions and random networks generated using Erd\H{o}s-R\'{e}nyi algorithm as well as using Watts-Strogatz random rewiring model forming the individual layers of the multiplex networks. Then, starting from both the layers depicted by the 1D lattice, we study how by rewiring the edges of one layer, that is, increasing structural randomness in one layer, spectral properties of the entire multiplex network get affected. (ii) Second, how the interplay of multiplexing strength and the network architecture of individual layer affects the spectra of the entire multiplex network. This investigation can give insight into dynamical behaviors or responses of real-world complex systems having multilayer network architecture as a consequence of change in the randomness in the structure of only one of its layers. Note that, here \textit{randomness} is ascribed to as random connections among the nodes.

\section{Model and Techniques}
A multilayer network can be defined as $A = (G, E)$ where $G$ = \{$g_{a}$ ; a = \{1,2,.....$m$\}\} is a set of graphs $g_{a}$= ($N_{a}$, $e_{a}$) for a = $1,2,....m$ layers of network and $E$ = \{ $e_{ab}$ ; a,b $\in$ \{1,2,.....m\}\}. If a=b, elements of $E$ are intra-layer connections i.e., connections between the nodes $N_a$ of the same layer $g_a$. For a$\neq$b, $E$ is the set of inter-layer connections between the nodes $N_{a}$ and $N_{b}$ of layer $g_{a}$ and $g_{b}$. The strength of connections between the layer $g_{a}$ and $g_{b}$ is denoted as $D_{x}$. There is a particular type of multilayer network called multiplex network, which consists of the same set of nodes in all the layers, which have different types of interactions. In such networks, inter-layer connections connect each node to its mirror node in the other layers.
Here, we consider bi-layer multiplex networks, with each layer having $N$ number of nodes. The adjacency matrix $A$ for such a two-layer multiplex network can be then written as;
\begin{align}\label{eq2}
A =  \left(
\begin{array}{cc}
\ A^1  & D_{x}I \\ 
D_{x}I & \ A^2 \\
\end{array}
\right)
\end{align}
where $ A^{1} $ and $ A^{2} $ are the adjacency matrices corresponding to layer $1$ and layer $2$, respectively, and $D_x$ is the multiplexing or inter-layer strength. The degree of $i^{th}$ node in layer $1$ is given by 
$ {k_{i}}^{1} $ = $\sum_{j=1}^{N}$ ${A_{ij}}^1$ + $1$ where ${A_{ij}}^1$ is the $ij$ entry of $A^{1}$ matrix and $1$ corresponds to the inter-layer connection for $D_x=1$. The 1D lattice network considered here has its each node connected to the same number of the neighboring nodes. The randomness in a network is introduced using the Watts-Strogatz SW model \cite{ws}. Starting with $N$ nodes where each node is connected with its $k$ nearest neighbors ($k/2$ each side), each connection of this 1D lattice network is rewired randomly with a probability $p_{r}$ resulting in $(p_{r}\times N \times k)/2$ long range connections among those nodes which were not connected in the original 1D lattice. On increasing the rewiring probability, for an intermediate value of $p_{r}$ the network undergoes the SW transition characterized by the high clustering coefficient and small characteristic path length. The clustering coefficient of a graph ($C$) is average of $C_i$ over all its vertices, which is defined as the ratio of actual edges between the neighbors of node $i$ to the possible number of edges between its neighbors, whereas the characteristic path length is the average of the shortest path between all the pairs of the nodes.
Additionally, we use the Erd\H{o}s-R\'{e}nyi (ER) model to generate random networks \cite{R1, R2, R3}. The ER model network of size $N$ is constructed by connecting every pair of nodes with a probability $p$. 

\begin{figure}[t]
\setlength{\abovecaptionskip}{6pt}
\setlength{\belowcaptionskip}{0pt}
\centering
    \includegraphics[width=0.8\textwidth]{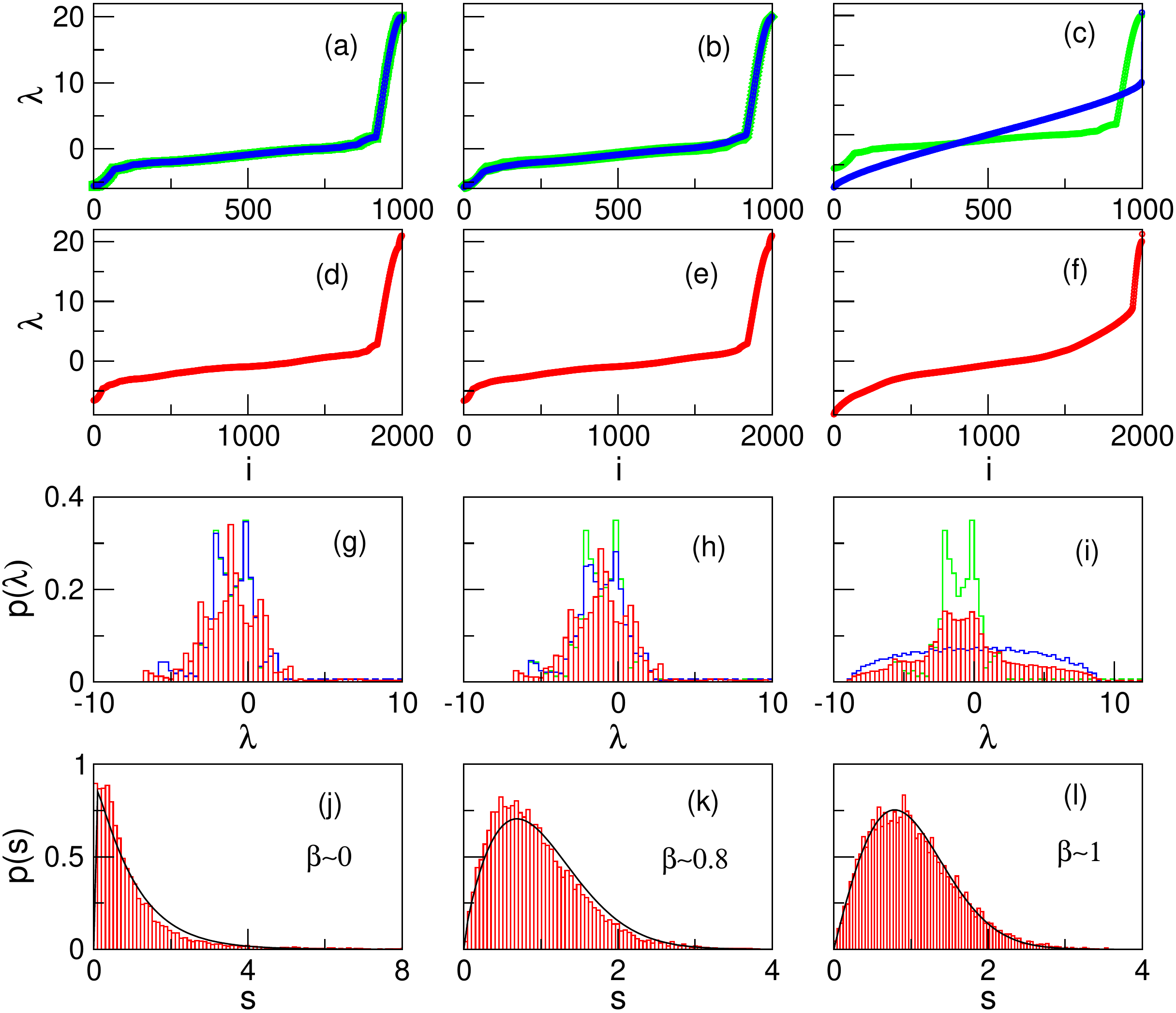}
    \caption{Spectral properties of 1D-rewired multiplex networks. Here, (a)-(c) Eigenvalues of the isolated networks, (d)-(f) eigenvalues of the multiplex network, (g)-(i) spectral density and (j)-(l) spacing distribution of the multiplex networks. Layer 1 is fixed as 1D lattice, and layer 2 represents the network at different values of the rewiring probability $p_{r}=0.0001, 0.01, 1$ from left to right. Each layer consists of $N_{1}=N_{2}=1000$ nodes and average degree $\langle k_1 \rangle=\langle k_2 \rangle=20$. Green color represents layer $1$, blue represents layer $2$ and red represents the multiplex network.}
\label{fig.2}
\end{figure}

The eigenvalues of an adjacency matrix $A$ of a multiplex network are denoted by $ \lambda_{i} $, $i = 1 , . . . , 2N$, then $2N$ becomes the size of the multiplex network and $\lambda_{1} < \lambda_{2} \le \lambda_{3} \le ........ < \lambda_{2N}$. To study the fluctuations of eigenvalues, it is customary in RMT to unfold the eigenvalues by a transformation given as;
\begin{equation}\label{eq3}
\bar{\lambda}_{i} = \bar{N}(\lambda_{i})
\end{equation}
where $\bar{N}$ is the averaged integrated eigenvalue density \cite{Mehta}. In the absence of an analytical form for the $\bar{N}$, the spectrum is unfolded numerically by the polynomial curve fitting. After unfolding the eigenvalues, the average spacings will be unity independent of the system. Using the unfolded spectra, we calculate the spacings as:
\begin{equation}\label{eq4}
 s_{i} =  \bar{\lambda}_{i+1} - \bar{\lambda}_{i} 
\end{equation}
NNSD is defined as the probability distribution [$P(s)$] of $s_{i}$. For the case of the Poisson statistics,
\begin{equation}\label{eq5}
P(s)= \exp(-s)
\end{equation} 
whereas for Wigner surmise,
\begin{equation}\label{eq6}
P(s) = \frac{\pi}{2}s\exp\left(- \frac{{\pi} s^2}{4}\right)
\end{equation}
Here, we would like to mention that though Wigner surmise was originally calculated for $N=2$ matrices which later turned out to be a good approximation for p(s) of size $N>2$ GOE matrices, there exists exact well-known formula of P(s) for GOE matrices \cite{Mehta}. For the intermediate cases, the spacing distribution is described by the Brody distribution\cite{RMT2} :
\begin{equation}\label{eq7}
 P^{\beta}(s)=A{s^{\beta}}\exp(-\alpha{s^{\beta + 1}})
\end{equation}                                 
where \begin{equation}\label{eq8}
A=(1+ \beta)\alpha
\end{equation}  and    \begin{equation}\label{eq9}
\alpha= \left[\Gamma\left(\frac{\beta+2}{\beta+1}\right)\right]^{\beta+1} 
\end{equation}
This is a semi-empirical formula characterized by the Brody parameter $\beta$. As $\beta$ goes from 0 to 1, the Brody distribution smoothly changes from the Poisson to the GOE statistics \cite{RMT2}. There exist many phenomenological models to study the NNSD indicating transition from Poisson-GOE transition or intermediate statistics. For instance, Berry-Robnik (BR) formula \cite{BR} has been applied to the level spacing distribution for a variety of shapes of quantum oval billiards \cite{BR}, also, validity of BR level spacing distribution has been demonstrated in a generic smooth plane billiard system \cite{BR}. Also, GOE-GUE transition depending on a parameter $\alpha$, for $\alpha=0$, GOE and for $\alpha=1$, the ensemble is Gaussian unitary \cite{O-E} and cross-over transition between Poisson-GOE-GUE \cite{crossover}. However, the Brody distribution has been widely studied, and there are well-known results for the case of single layer networks to measure the transition/mixture of GOE and Poisson statistics and for real-world networks also \cite{sj1, sj2, RMT_geometric, RMT_powergrid, RMT_RGG, RMT_Normal_mode, RMT_brain, RMT_skeleton, RMT_quantum, RMT_random, RMT_ADHD, RMT_ginibre, RMT_limb, RMT_functional}. Hence we use Brody distribution in the current article.

To calculate $p(s)$, eigenvalues from both the ends are removed and only the smooth bulk part of the spectra is considered. Also, any degeneracy arising in the eigenvalues is removed. Thereafter, the unfolding process is performed by numerical polynomial curve fitting of degree $n$. An ensemble of $10$ random realizations of the underlying network structure is considered, and the unfolding of the eigenvalues as well as calculations of the spacing are done separately for each realization. Thereafter, the combined spacings for all the realization for a particular network structure is fit by the Brody distribution $P_{\beta}(s)$. This fitting gives an estimation of $\beta$ and consequently identifies whether the spacing distribution of a given network is Poisson, GOE, or an intermediate of these two. 

\begin{figure}[t]
\setlength{\abovecaptionskip}{6pt}
\setlength{\belowcaptionskip}{0pt}
\centering
    \includegraphics[width=0.8\textwidth]{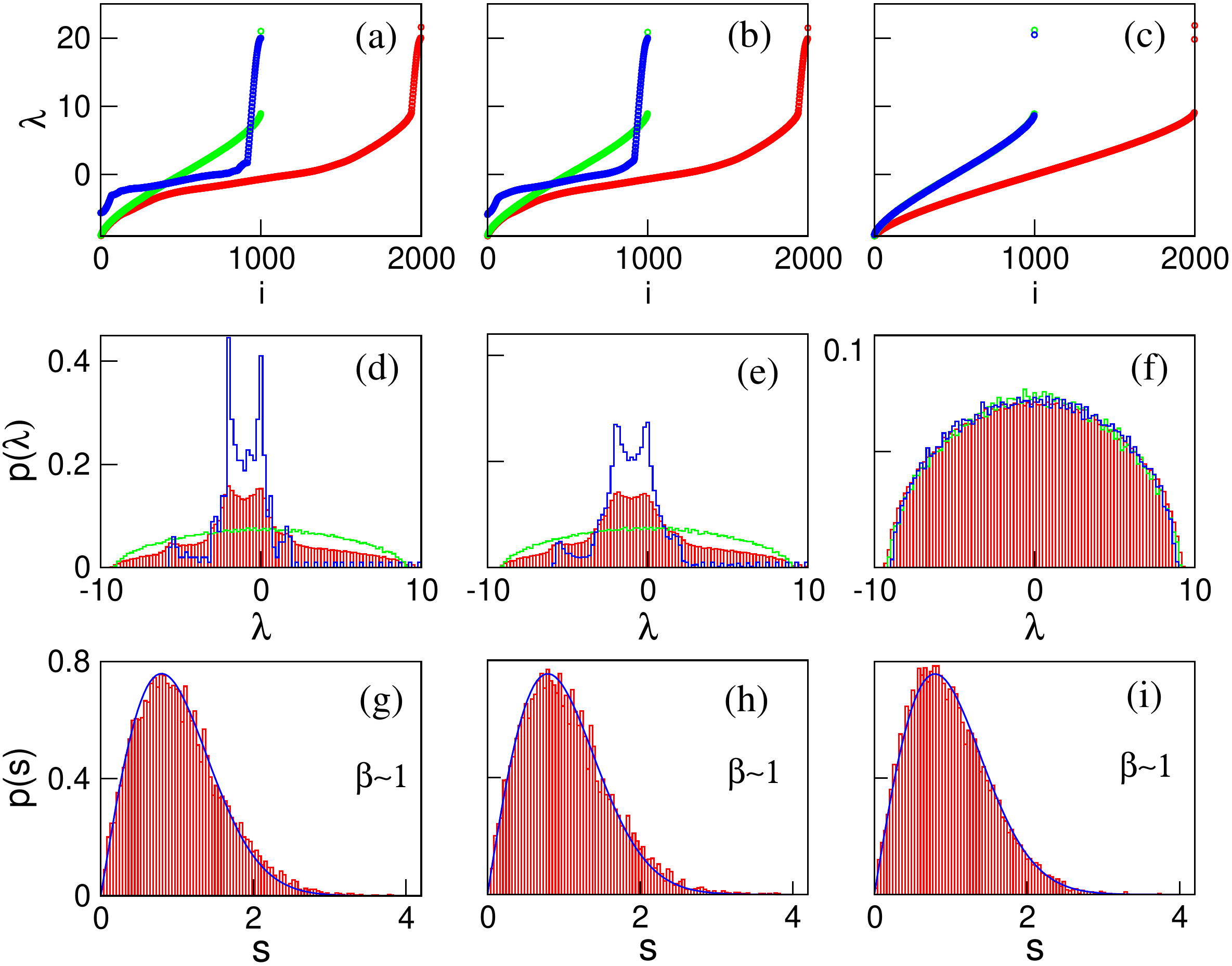}
    \caption{Spectral properties of ER-rewired multiplex networks. Here, (a)-(c) Eigenvalues, (d)-(f) spectral density and (g)-(i) spacing distribution of the multiplex networks. Layer 1 is fixed as ER random network, layer 2 consists of SW networks generated at different rewiring probabilities $p_{r}=0, 0.01, 1$ from left to right. Each layer consists of $N_{1}=N_{2}=1000$ nodes and average degree $\langle k_1 \rangle=\langle k_2 \rangle=20$. Green color represents layer $1$, blue represents layer $2$ and red represents the multiplex network.}
\label{fig.3}
\end{figure}

\section{Results and discussion}
We investigate the spectral properties of bi-layer multiplex networks under the RMT framework. Through the spacing distribution of the eigenvalues, we demonstrate that how RMT analysis of spectra of the entire multiplex network can capture or get affected by the {\it randomness} of the individual layer.
Additionally, by varying the strength of the inter-layer connections, i.e., the multiplexing strength, whether the effect of rewiring in one layer on the spectra of the entire multiplex networks can be contained. 
We first present the numerical results for multiplex networks consisting of one layer represented by the 1D lattice or ER random network and the second layer constructed using the Watts-Strogatz SW algorithm for different values of the rewiring probabilities.
The identical network structure in both the layers yields the spectral density same as that of the single layer network of the same size and topology. Note that identical structure in both the layers means different realizations of the network by keeping overall structural properties same, i.e., 1D, ER random, SW with the same rewiring probability. However, for both the layers of a multiplex network having different network architectures, the spectral density of the entire multiplex network can be more complex bearing resemblance with the shape of spectral densities of the isolated layers. For example, if one layer is represented by the ER random network and the other layer has 1D network architecture, the spectral density of the multiplex network consisting of these layer displays the semicircular distribution corresponding to the spectra of the isolated ER random network along with few multi peaks corresponding to the spectra of the isolated 1D regular network.
Also, it is known that NNSD of 1D lattice with a very small number of rewirings ($p_r$ being very small) follows Poisson statistics, whereas that of the ER random networks follow GOE. Here, we show that on multiplexing a 1D regular network with another 1D regular network, NNSD of the resulting multiplex network follows Poisson statistics irrespective of multiplexing strength. Whereas, NNSD of the multiplex networks with ER random network forming both the layers is superposition of $2$ GOEs for low multiplexing with a transition to GOE statistics as multiplexing strength $(D_x)$ is increased.

\subsection{Effect of randomness in one layer}

Next, we systematically present detailed results for the effect of {\it randomness}, i.e., random rewiring to an initial 1D lattice structure on the spectral properties of the entire multiplex network.
The random rewiring of an isolated 1D lattice even with a very small probability $p_{r}$ has been shown to lead to drastic changes in the Brody parameter \cite{sj1}. Similarly, for a multiplex network consisting of both the layers represented by the 1D regular networks, rewiring with a small $p_{r}$ value in one of the layers may lead to a drastic changes in the spectral properties of the entire multiplex network.
For example, Fig.2 presents results for a multiplex networks consisting of one of the layers (say layer 2) randomly rewired using the Watts-Strogatz algorithm and another layer represented by a 1D lattice. In the following first, we present results about the impact of $p_r$ which accounts for {\it randomness} in one layer on the spectra of entire multiplex networks.


\begin{figure}[t]
\setlength{\abovecaptionskip}{6pt}
\setlength{\belowcaptionskip}{0pt}
\centering
    \includegraphics[width=0.8\textwidth]{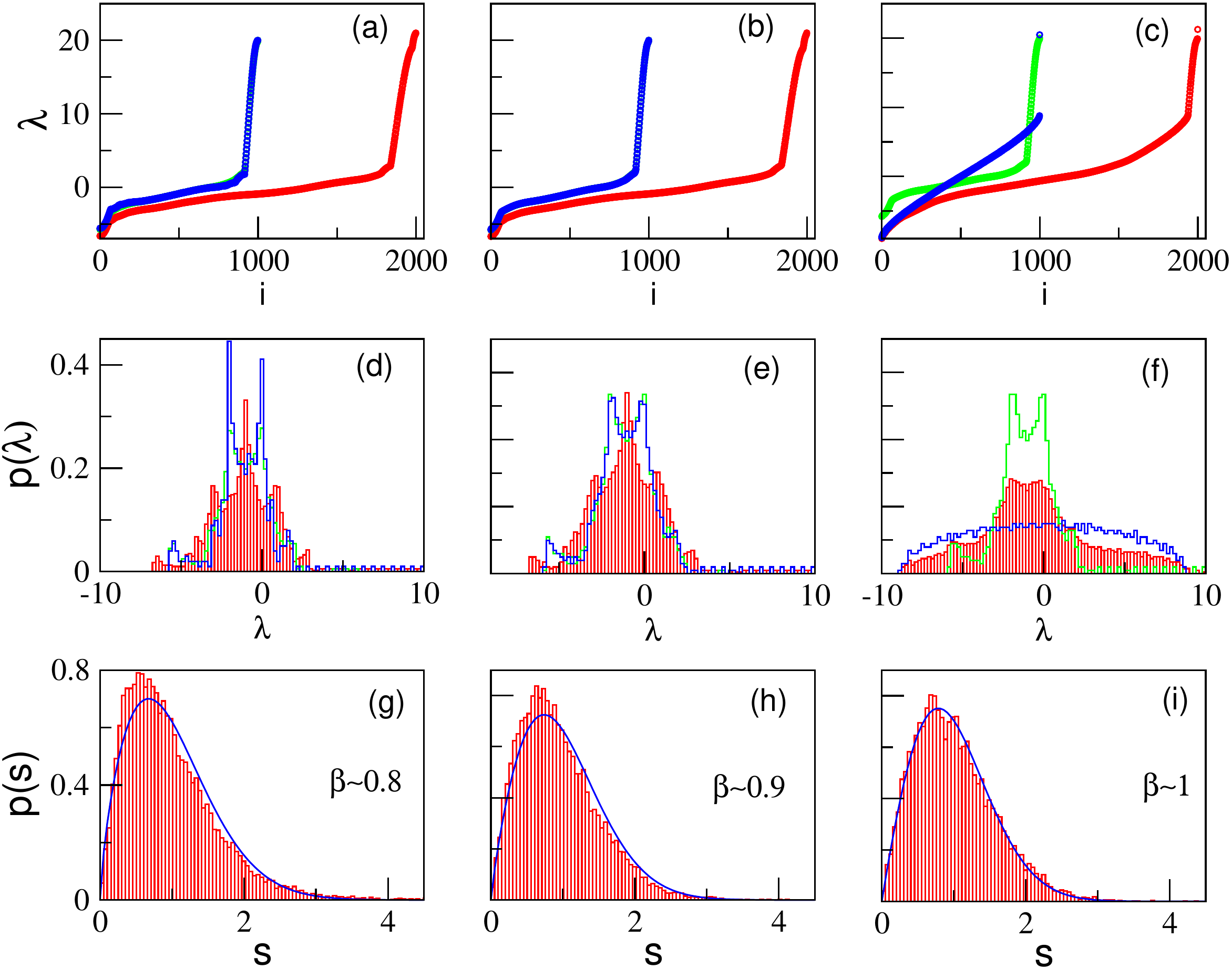}
    \caption{Spectral properties of small-world rewired network. Here, (a)-(c) Eigenvalues, (d)-(f) spectral density and (g)-(i) spacing distribution of multiplex network. Layer 1 is fixed as small-world network, in layer 2 network is rewired with the rewiring probability $p_{r}=0, 0.01, 1$ from left to right. Each layer consist of $N_{1}=N_{2}=1000$ and average degree $\langle k_1 \rangle=\langle k_2 \rangle=20$. Green color represents layer $1$, blue represents layer $2$ and red represents the multiplex network.}
\label{fig.4} 
\end{figure}

\subsubsection{Spectra of 1D-rewired multiplex network:}
Fig.~\ref{fig.2} presents the spectral density and the spacing distribution of the multiplex networks for various values of $p_r$. To understand the variations in the spectra and the spacing distribution with the change in the rewiring probability $p_r$ for layer 2, the spectra of the individual isolated layers are plotted along with that of the multiplex networks consisting of these layers. For small $p_r$ values, as expected the eigenvalues of the individual layer keep coinciding with each other. However, for a large variation in $p_r$, which leads to a large deviation in one of the layers from the initial 1D regular structure, the eigenvalues of the entire multiplex network also change. Note that here the network parameters, i.e., size and the average degree, remains the same, and only the network structure changes as a consequence of the random rewiring. It is also worth noting here that the random rewiring performed using the Watts-Strogatz algorithm does not bring significant changes in the degree distribution from that of the initial 1D lattice.
Fig.~\ref{fig.2}(a)-(c) and Fig.~\ref{fig.2}(d)-(f) depict the eigenvalues of the isolated layers and the corresponding multiplex networks, respectively. On rewiring the 1D lattice network forming one of the layers, as the rewiring probability increases the spectral distribution of that layer which has multi-peak complex structure starts broadening and acquires a semi-circular shape at $p_{r}=1$ (Fig.~\ref{fig.2}(g)-(i)). The same trend is followed by the spectral density of the entire multiplex network; initially, for $p_r=0$, the spectral density is same as that of the 1D lattice network. As $p_{r}$ increases, due to the broadening in spectral density of the second layer, spectral density of whole multiplex network starts broadening. At $p_{r}=1$, the spectral distribution of the multiplex network acquires a semicircular shape consisting of multi peaks. Fig.~\ref{fig.2}(j) presents the spacing distribution of a bi-layer network with both layers represented by the 1D lattices \cite{CLR}. The spacing distribution clearly follows the Poisson statistics ($\beta0$). By fitting the spacing distribution corresponding to different $p_{r}$ values with the Brody formula, we estimate the value of the Brody parameter $\beta$ for the various different rewiring probabilities. These values indicate that on keeping one layer fixed as 1D lattice, as we rewire the other layer, initially for small values of $p_r$, the spectra of the combined multiplex network keeps following Poisson statistics as also followed by the isolated layer. However, as $p_{r}$ increases, $\beta$ increases and reaches very close to one reflecting the GOE statistics even for very small value of $p_r$ (Fig. 2(k)). It shows that a very small random rewiring in only one layer (here layer 2) can lead to diminishing the regularity of the entire multiplex network leading to short range correlation in the eigenvalues.

\begin{figure}[t]
\setlength{\abovecaptionskip}{6pt}
\setlength{\belowcaptionskip}{0pt}
\centering
    \includegraphics[width=0.7\textwidth]{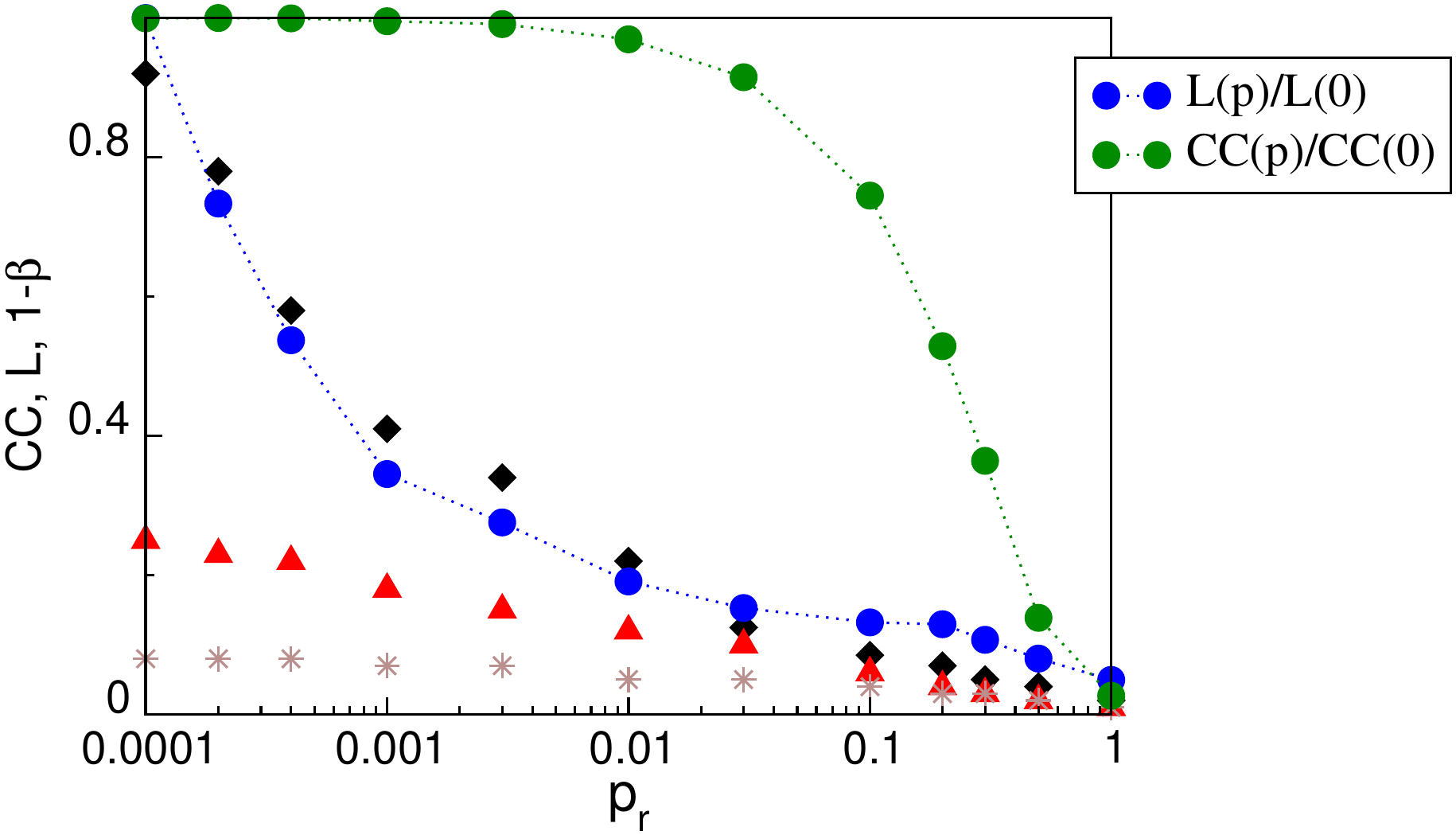}
    \caption{The shifted Brody parameter $1-\beta$ of the three multiplex networks (with one layer fixed as 1D, ER random and SW network while the second layer undergoes transition from 1D lattice to random network) is compared with the two network parameters, normalized characteristic path length and normalized clustering coefficient of a rewired network. Each layer consists of $N_{1}=N_{2}=1000$ nodes with average degree $\langle k_1 \rangle=\langle k_2 \rangle=20$. The data are averaged over 20 random realizations of the rewiring process for each value of $p_{r}$. The $ \color{black} \blacklozenge$ , $ \color{red} \blacktriangle$ and $ \color{Orchid} \ast$ represent 1D lattice, SW and ER network fixed in one layer respectively, while the other layer undergoes rewiring.}
\label{fig.5}
\end{figure} 

\subsubsection{Spectra of Random- rewired multiplex network:}
Here, we present results for multiplex networks consisting of one layer (say layer 1) as the ER random network and another layer (layer 2) as 1D regular network. Then we keep the first layer fixed to the ER random structure and rewire the second layer using the Watts-Strogatz model with the probability $p_r$, which changes from 0 to 1. We investigate the spectral properties of the entire multiplex networks for the entire range of $p_r$. Fig.~\ref{fig.3} presents the spectral density and the spacing distribution of such multiplex networks along with those of the individual layers in isolation. 
A very small change in $p_r$ does not bring much changes in the spectral properties of the multiplex networks (see Fig. 3) as the multiplex network in this setup already consists of one layer, which is completely random. 
Note that the largest and the second-largest eigenvalues will always be separated from the bulk part of the eigenvalues for a multiplex network having two layers as for a network consisting of $N$ nodes and $N_c$ communities, $N_c$ eigenvalues are separated from bulk (N-$N_c$) part of the eigenvalues \cite{community}. For the bi-layer multiplex networks, there will be always two clear communities (corresponding to two layers) irrespective of absence or presence of further community structure in the individual layer. In a random-rewired multiplex network, initially for $p_r=0$ the spectral density exhibits a combination of both the semicircular and the multipeak shape (Fig.~\ref{fig.3}(d)) owing to ER random and 1D lattice network architectures of the individual layers. As rewiring probability $p_{r}$ is increased (Fig.~\ref{fig.3})(e)-(f), the multipeak density is redistributed and merged with the semicircular structure. Due to the presence of ER random network in one of the layers of the multiplex network, there exists already a spread of {\it randomness} in the entire multiplex network reflected in the spacing distribution which follows GOE statistics even for another layer being completely regular ($p_r=0$). An increase in the {\it randomness} of this layer ($p_r > 0)$ keeps yielding the GOE spacing distribution (Fig.~\ref{fig.3}(g)-(i)).
As expected, one layer being fixed to an ER random network, irrespective of structure of the second layer, the NNSD of the entire multiplex network follow GOE. One layer of multiplex networks being random is enough to introduce short range correlations in the eigenvalues of the entire multiplex networks.

\subsubsection{Spectra of SW-rewired multiplex network:} \label{subsection}
The small-world networks at the rewiring probability close to the SW transition are neither wholly regular nor random. They are characterized by the high clustering coefficient, which is close to that of the 1D regular networks, and low characteristic path length, close to that of the ER random networks. Here we present results for multiplex networks consisting of one layer as a SW network with $p_r=0.01$, which is kept fixed throughout. The other layer is initially considered as 1D lattice network, and subsequently, the random rewiring is performed for different values of $p_r$. The NNSD of isolated SW networks have been shown to follow GOE statistics \cite{sj1, sj2, cam}. Here, we show that the multiplexing of a SW network with any network keeps the sufficient {\it randomness} in the entire multiplex networks so that there exists short range correlations in the eigenvalues yielding GOE statistics for NNSD.
Fig.~\ref{fig.4}(a)-(c) plots the eigenvalues of the individual layers as well as that of the multiplex networks consisting of these layers. Fig.~\ref{fig.4}(d)-(f) depicts spectral density of the isolated layers as well as that of the multiplex network. Here, one layer is fixed to the SW network with $p_r=0.01$. On rewiring the other layer harbouring a 1D lattice network, as $p_r$ increases, the density distribution starts broadening from an initial multi-peak complex shape and attains the typical semicircular structure of random networks at $p_{r}=1$. The spectral distribution broadens to the semicircular shape as $p_{r}$ tends to $1$. Fig.~\ref{fig.4}(g)-(i) plots spacing distribution of a multiplex network (with one layer fixed as SW network) and other with different values of the rewiring probability.

Fig.~\ref{fig.5} presents the change in the Brody parameter $\beta$ for a multiplex network as $p_{r}$ changes from $0$ to $1$ and compares it with two structural parameters, clustering coefficient $C(p_{r})$ and characteristic path length $L(p_{r})$. We have normalized $L$ and $C$ by their respective values of 1D lattice, i.e., $L(0)$ and $C(0)$. Fig.~\ref{fig.5} plots $1-\beta$ with the normalized $L$ and $C$ for three different multiplex networks. The random rewiring in the single layer network has been shown to demonstrate the SW transition and Poisson-GOE transition occurring at the same value of the rewiring probability \cite{sj1}. For a multiplex network with one layer fixed as a 1D lattice and the other layer being rewired with a probability $p_r$, the similar behavior is observed as that of a single layer. For multiplex networks consisting of 1D lattice fixed in one layer, the Poisson to the GOE transition occurs at the same rewiring probability $p_{r}$ at which the SW transition occurs. It reflects that after multiplexing any network with a 1D lattice network, the spectral statistics of a multiplex network remains the same as that of the single layer network before multiplexing. However, on fixing one layer as ER random network and rewiring other layer with different values of $p_r$, as $p_{r}$ varies from $0$ to $1$, the entire multiplex network retains enough {\it randomness} throughout the change of the rewiring probability so that there keeps existing the short range correlations in the eigenvalues and the spacing distribution keeps following the GOE statistics for all the $p_r$ values. Similarly, if one layer of a multiplex network is represented by a SW network, there will always be some amount of randomness present in the multiplex network which increases on increasing $p_r$ to $1$.

\begin{figure}[t]
\setlength{\abovecaptionskip}{6pt}
\setlength{\belowcaptionskip}{0pt}
\centering
    \includegraphics[width=0.8\columnwidth]{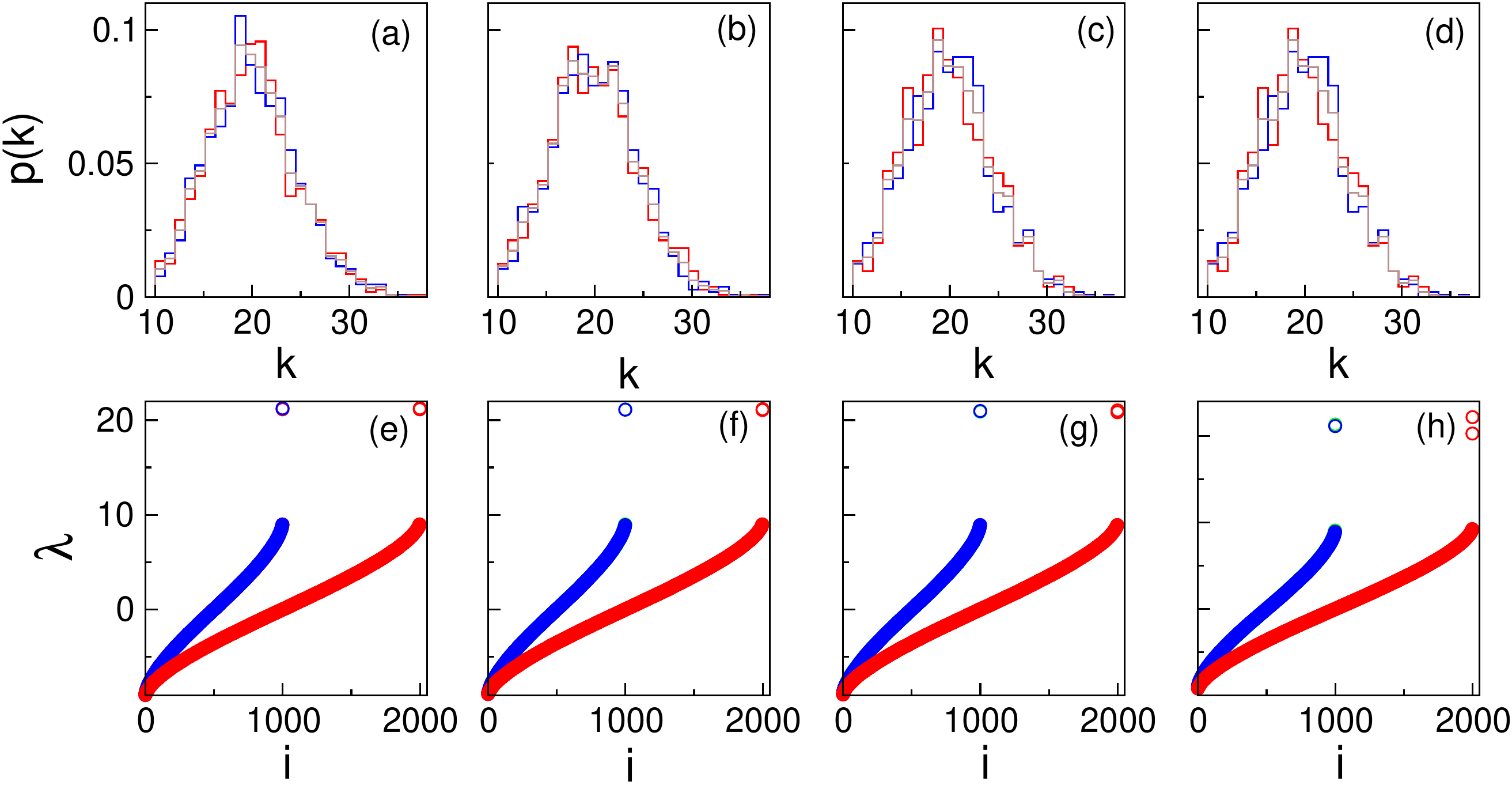}
    \includegraphics[width=0.8\columnwidth]{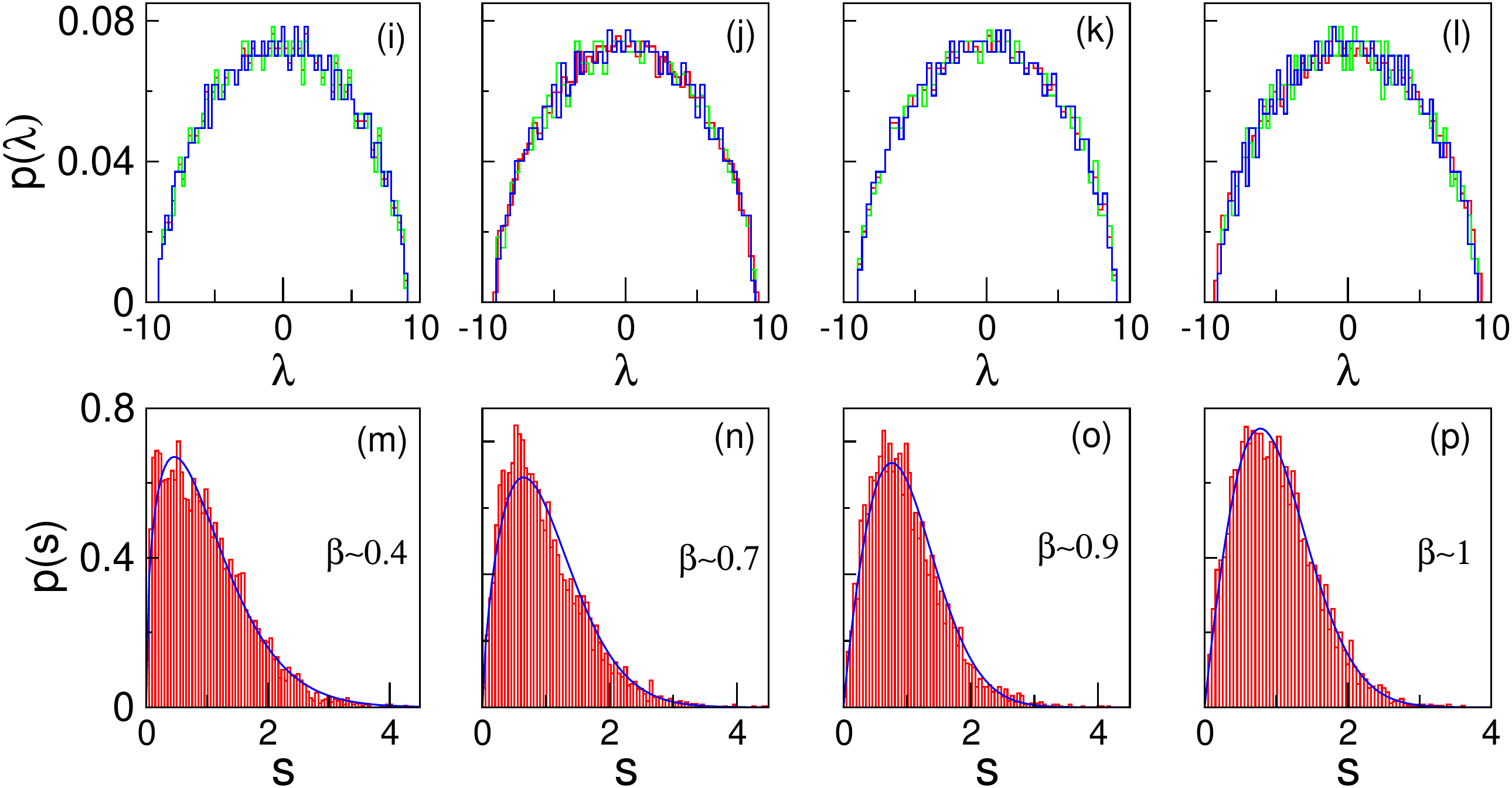}
    \caption{Spectral properties of the ER-ER random multiplex networks. Here, (a)-(d) Degree distribution, (e)-(h) eigenvalues, (i)-(l) spectral density and (m)-(p) spacing distribution  of isolated networks in both the layers and multiplex network. $D_x= 0.01, 0.05, 0.1, 1$ (left to right). All results shown here are for bi-layer networks with $N_{1}=N_{2}=1000$ and average degree $\langle k_1 \rangle=\langle k_2 \rangle=20$. The green, blue and red color curve correspond to the eigenvalues (green and blue coinciding) (e)-(h) and spectral density (i)-(l) of ER network in layer $1, 2$ and of multiplex network respectively.}
\label{fig.6}
\end{figure}

\subsection{Role of multiplexing strength}
Next, we investigate the impact of multiplexing strength on the relationships between structural properties of one layer and spectral properties of the multilayer network.
Particularly, we focus on how the change in the spectral properties of the entire multilayer network arising due to change in the {\it randomness} in one or both of its layers varies or is contained by changing the multiplexing strength $D_x$.

\begin{figure}[t]
\setlength{\abovecaptionskip}{6pt}
\setlength{\belowcaptionskip}{0pt}
\centering
    \includegraphics[width=0.7\textwidth]{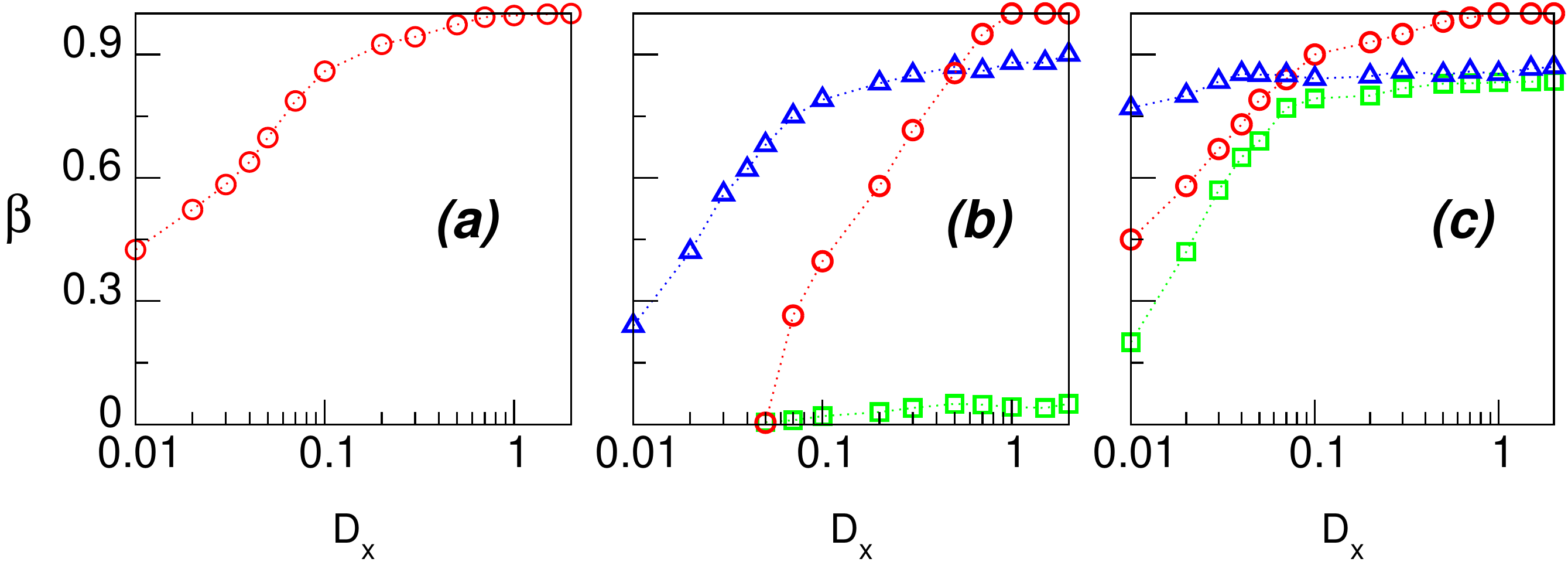}
    \caption{Brody parameter $\beta$ is plotted with  $D_x$ on the logarithmic scale. (a) Both layers are fixed as ER random networks. (b) One layer is fixed as 1D lattice network and other layer is rewired. (c) One layer is fixed as SW network and other layer is rewired. $D_x$ is varied from $0.01$ to $2$. All results shown here are for bi-layer networks with $N_{1}=N_{2}=1000$ and average degree $\langle k_1 \rangle=\langle k_2 \rangle=20$. The $\color{green} \square$, $\color{blue} \triangle$ and  $\color{red} \circ$ correspond to the network rewired at $p_r=0, 0.01, 1$ respectively in second layer.}
\label{fig.7}
\end{figure}

\begin{figure}[!htb]
    \centering
    
    { \includegraphics[width=0.5\linewidth]{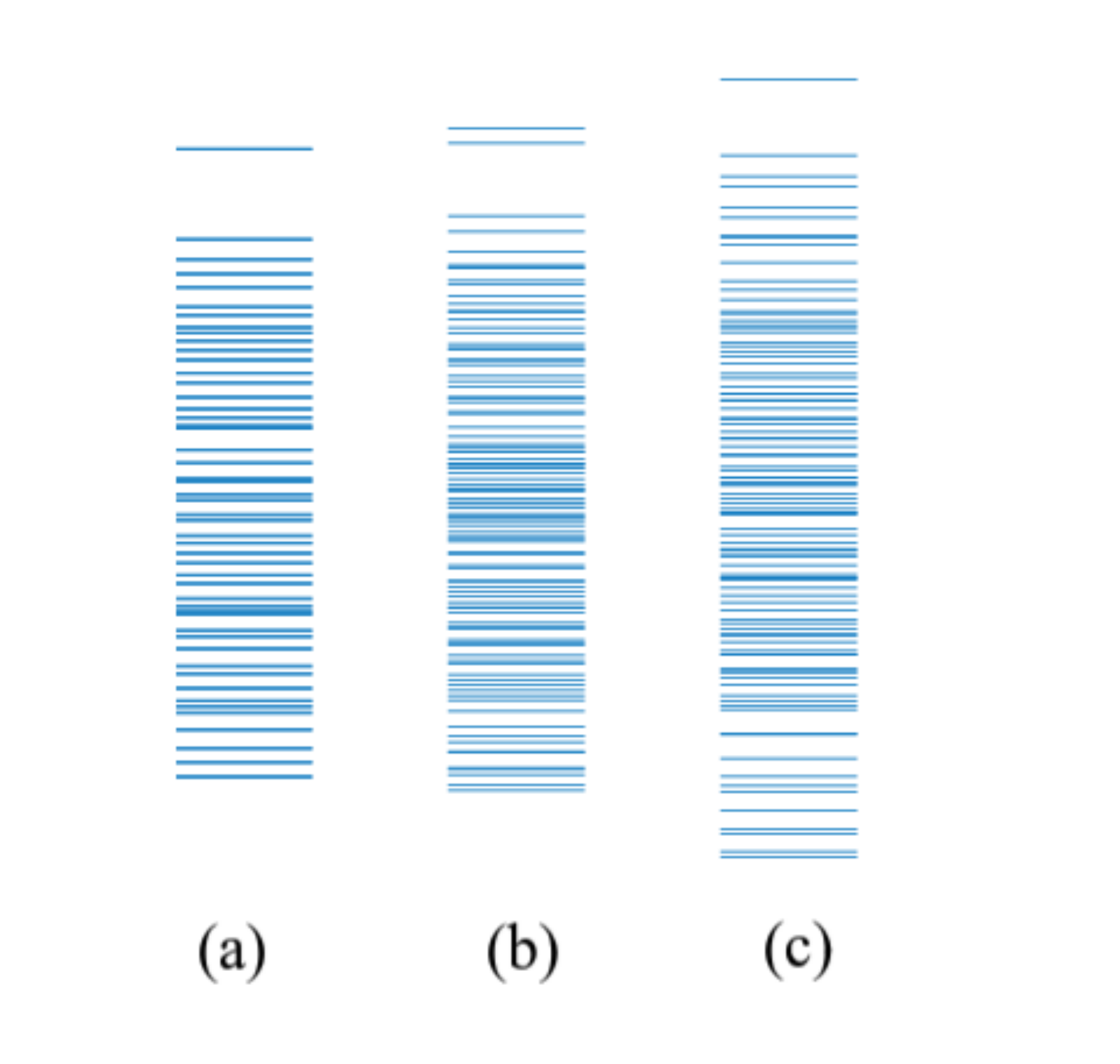}}  
    {\includegraphics[width=0.4\linewidth]{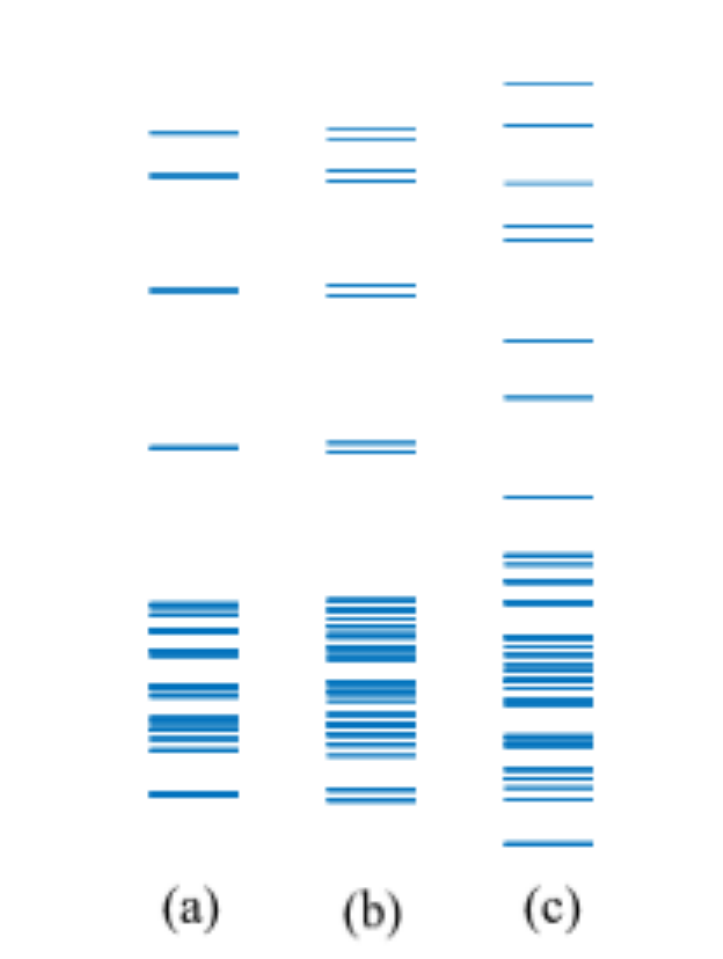}}
 
    \caption{(Color online) Spectra of multilayer networks with different values of the inter-layer coupling strength. (a), (b), (c) are  for $D_x=0.01, 0.1$ and $1$, respectively of a ER-ER multiplex network (left) and 1D-1D lattice (right). Here, for visual clarity $N_{1}=N_{2}=100$ in each layer and average degree $\langle k_1 \rangle=\langle k_2 \rangle=20$.}
\label{fig.8}
\end{figure}  


First, we consider the ER random network forming both the layers of the multiplex networks. Due to the identical networks in both the layers, the entire multiplex network demonstrates the same spectral statistics as depicted by any of its isolated layers. 

\subsubsection{Random-random multiplex network:}
Fig.~\ref{fig.6} presents various spectral properties of multiplex networks consisting of ER random networks in both the layers as a function of the multiplexing strength. These results are presented for ER random networks of the same average degree in both the layers (Fig.~\ref{fig.6}(a)-(d)). Also, the spectral density of a multiplex network consisting of the ER network in both the layers exhibit semi-circular distribution similar to that exhibited by the isolated ER random networks (Fig.~\ref{fig.6}(i)-(l)). The two largest eigenvalues of this multiplex network which for very small $D_x$ values are almost the same as the largest eigenvalue of the isolated layer and hence keeps coinciding (Fig.~\ref{fig.6}(e)-(g)). With an increase in $D_x$, the two largest eigenvalues get separated from each other (Fig.~\ref{fig.6}(h)). Fig.~\ref{fig.6}(m)-(p) plots NNSD for the multiplex networks for various different inter-layer coupling strength depicting the change in the short range correlations in the eigenvalues with the change in the multiplexing strength. For very small values of $D_x$, the spacing distribution is dominated by the small spacings which transits through intermediate to GOE statistics as $D_x$ approaches $1$. This is also shown by Fig.~\ref{fig.7}(a) which plots the Brody parameter as a function of $D_x$ for ER random-random multiplex networks. At smaller values of $D_x$, the spacing distribution is superposition of 2 GOEs. With an increase in the multiplexing strength $D_x$, the Brody parameter $\beta$ increases to $1$ with the spacing distribution following the GOE statistics. 
These results drawn from simulations demonstrate that even for both the layers of multiplex networks being random, the impact of this {\it randomness} on the spectral properties can be controlled by changing the coupling strength between the layers. To illustrate this further, we plot spectral lines (Fig.~\ref{fig.8} left) for rather a small size network which shows the impact of a change in the inter-layer coupling on the eigenvalues. The spectral lines which were coinciding for smaller $D_x$ values, being not strong enough to perturb the eigenvalues of the individual layer (Fig.~\ref{fig.8}(a) left), get separated as $D_x$ is increased, thus giving rise to an increase in the spacings between the consecutive eigenvalues leading to the transition to the GOE statistics. 

\begin{figure}[t]
\setlength{\abovecaptionskip}{6pt}
\setlength{\belowcaptionskip}{0pt}
\centering
    \includegraphics[width=0.7\textwidth]{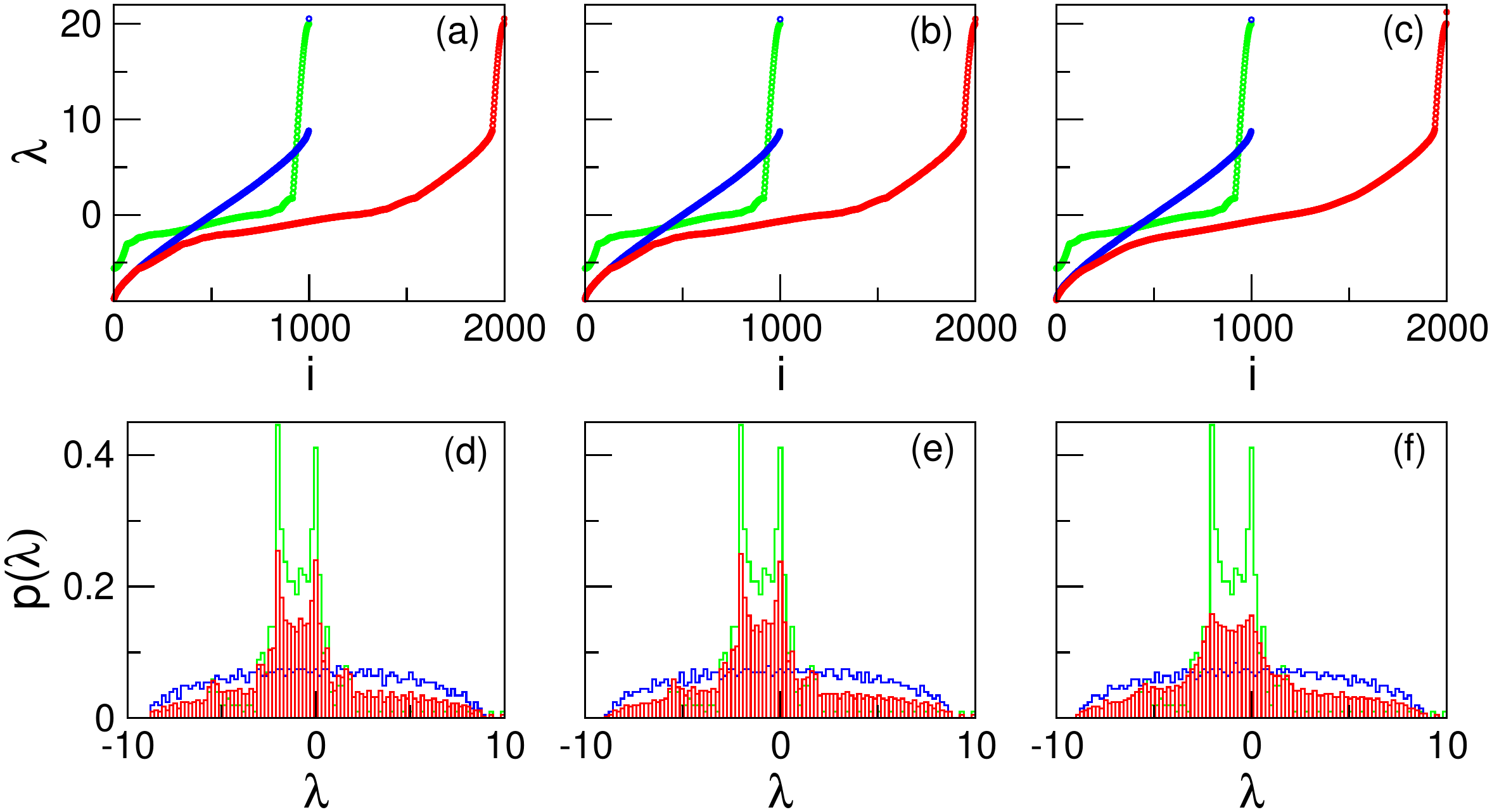}
    \caption{Spectral properties of a 1D lattice-ER random multiplex network. Here, (a)-(c) Eigenvalues and (d)-(f) spectral density of the isolated networks forming both the layers along with the multiplex network for $D_x$ = $0.01, 0.1, 1$ (left to right). Green corresponds to layer $1$ (1D lattice), Blue corresponds to layer $2$ (Random) and red corresponds to multiplex network. Each layer consist of $N_{1}=N_{2}=1000$ nodes and average degree $\langle k_1 \rangle=\langle k_2 \rangle=20$.  }
\label{fig.9}
\end{figure}
  
The earlier sections have shown that multiplexing a SW network with a 1D regular network does not affect the short range correlations in the eigenvalues or {\it randomness} of a network. Upon multiplexing it with a random network leads to an increase in the {\it randomness} of the entire multiplex network, which is not surprising. However, irrespective of the network architecture in the other layer, the spacing distribution keeps exhibiting a close resemblance ($\beta$ values remain very close) for differing values of the rewiring probability. It raises a question that if one of the layers has the SW network architecture, the random rewiring of the edges in other layer does not affect the spectral statistics (short range correlations) of the entire multiplex network, can it be affected by change in the multiplexing parameter $D_x$? 
Fig.\ref{fig.7}(c) plots $\beta$ as a function of $D_x$ for SW-rewired multiplex networks at different values of the rewiring probability. It shows that for a multiplex network consisting of SW networks in both the layers ($p_{r}=0.01$ for the rewired layer), an increase in the  inter-layer coupling strength does not have significant impact on the spacing distribution of the spectra of the entire multiplex network, and the Brody parameter keeps fluctuating around a fixed value.
Fig.\ref{fig.7}(c) also plots $D_x$ versus $\beta$ for the rewired-1D multiplex network ($p_{r}=0$), and rewired-ER multiplex network ($p_{r}=1$). Here, it can be seen that a decrease in the value of $D_x$ leads to a decrease in the $\beta$ value, i.e. {\it randomness} of the entire multiplex network decreases under the RMT framework providing to us a more clear picture of the role of the multiplexing strength. Additionally, with a decrease in $D_x$, {\it randomness} of the multiplex network consisting of SW network in one of the layers is decreased irrespective of the other layer being random or 1D lattice. However, as discussed in the earlier section (subsection \ref{subsection}) that irrespective of the network architecture of the second layer (Random, 1D lattice or network with any value of the rewiring probability), if one of the layers is represented by the SW network, the spacing distribution of the entire multiplex network remains almost the same with a change in the network architecture of the other layer. Therefore, the multiplexing parameter can be seen as a way to control the influence of one layer on the entire multiplex network.

\subsubsection{1D lattice - rewired multiplex networks }
This section shows that the random rewiring of the edges with a small probability, which corresponds to a small perturbation in the initial network structure of one layer, is enough to affect the spectra of the entire multiplex network in terms of its nearest neighbor spacing distribution. Further, this impact of changes in the structural properties in one layer on the entire multiplex network can be controlled by changing the multiplexing strength.

Fig.~\ref{fig.7}(b) plots $\beta$ as a function of the multiplexing strength for a multiplex network with the 1D lattice network representing one layer and a rewired network representing the other layer with $p_{r}=0.0001$. For all the $D_x$ values, the spacing distribution follows the Poisson statistics. Fig.~\ref{fig.8}(right) presents spectral lines of the multiplex networks consisting of 1D lattice network in both the layers for different values of the inter-layer connection strength. Upon multiplexing, the behavior of eigenvalues remains almost the same, they do not display any noticeable spread on increasing the multiplexing strength, and the spacing distribution keeps following the Poisson statistics throughout the $D_x$ values. Here, instead of 1D lattice network in both the layers, 1D lattice in one layer is taken with the rewired network at a very small probability $p_{r}=0.0001$ to avoid degenerate eigenvalues \cite{Dup_nodes1}. Fig.~\ref{fig.7}(b) plots the Brody parameter for layer 1 being represented by 1D lattice and layer two with the rewired network for two different values of $p_{r}= 0.01$ and $1$. Any change in the spectral statistics of the multiplex network arising due to the change in the rewiring probability of one layer varies with the variation in the multiplexing strength. Fig.~\ref{fig.9} presents the eigenvalues and spectral density of the multiplex network with 1D lattice forming the layer 1 and random network ($p_r=1$) forming the layer 2 for different values of $D_x$. As $D_x$ increases, the peak near $\lambda=0$ decreases (Fig.~\ref{fig.9}(d)-(f)) accompanied with a redistribution of the eigenvalues and changes in the spacing distribution measured by $\beta$.

\subsection{Layer with block subnetwork structure:}

In many real-world networks such as social \cite{Social1} and web networks \cite{web1}, nodes tend to divide into groups, with dense connections within the groups and sparse connections between the groups. Such types of structures can be modeled using networks having block subnetworks with underlying adjacency matrices having block structure \cite{block, block1}. Here, we consider the block model to construct one of the layers of a multiplex network in the following manner\cite{block, block1}. Each block of the adjacency matrix represents a community and off-diagonal blocks correspond to the links between communities. We present results for multiplex networks consisting of  one layer being represented by 1D lattice, and the other layer consisting of 2 subnetworks with the corresponding matrix being a 2-block diagonal matrix. We introduce random connections between these two subnetworks with a probability $q$ leading to the matrix having 2-block matrix with the probability $p$ and $q$ being the intra-block and inter-block connections, respectively. We first present the results of the multiplex network with the first layer formed by a 1D lattice and the second layer formed by a network of two blocks, one being ER random and the other 1D lattice with $q=0$. For such an architecture, the spacing distribution is intermediate between the Poisson and the GOE statistics. However, on introducing connections between the blocks (1D and ER) with a very small probability, the NNSD of the entire network follows GOE statistics.


\begin{figure}[t]
\setlength{\abovecaptionskip}{6pt}
\setlength{\belowcaptionskip}{0pt}
\centering
    \includegraphics[width=0.7\textwidth]{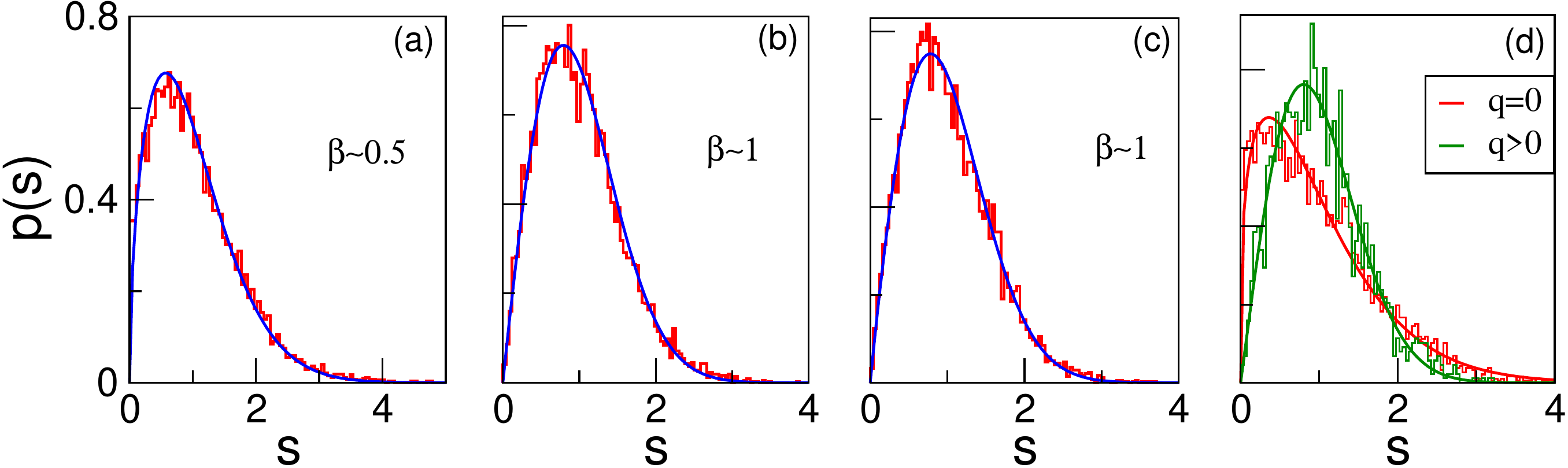}
    
    \vspace{0.3cm}
    
    \includegraphics[width=0.4\textwidth]{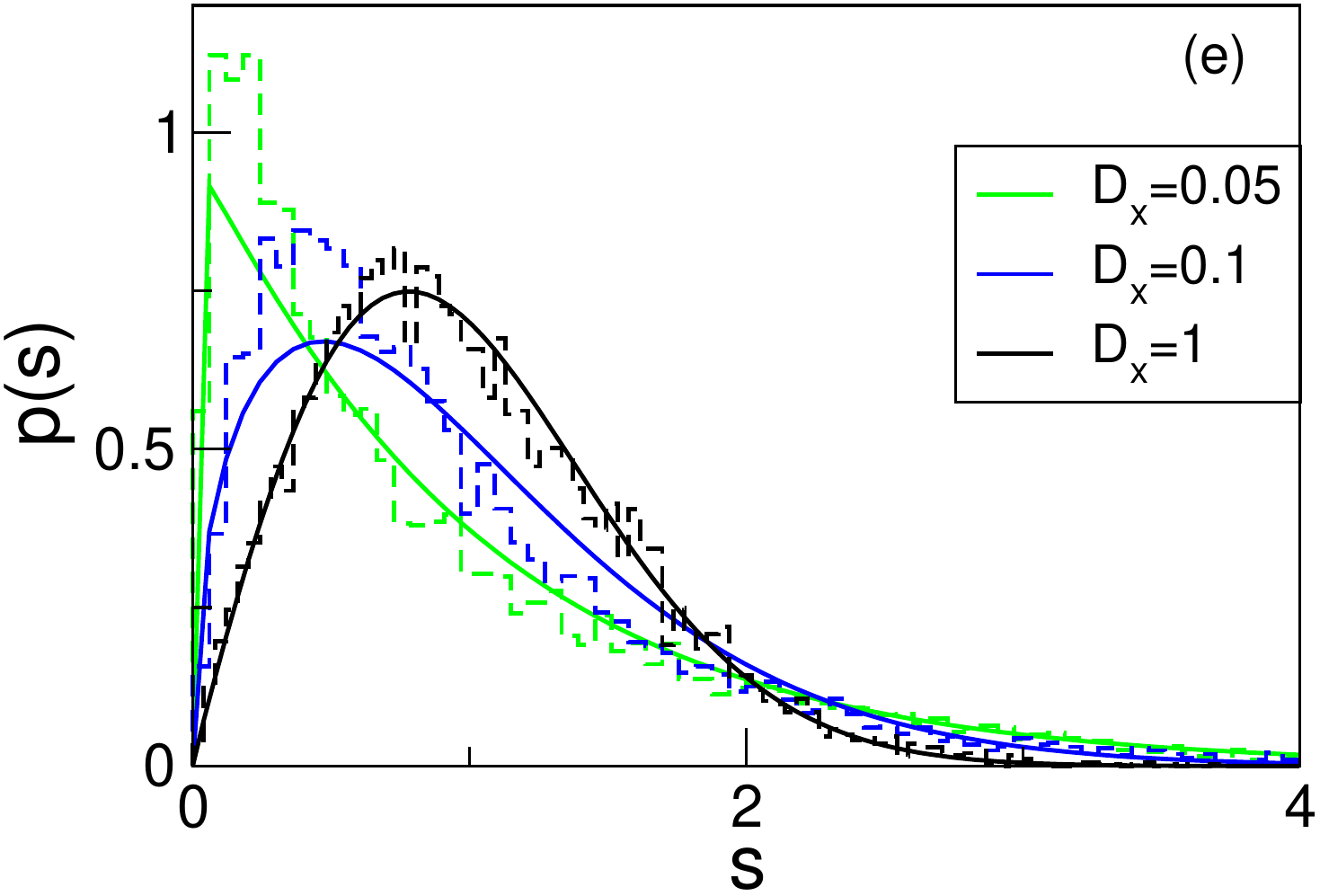}
    \caption{Spacing distribution of multiplex networks. (a) When in layer $2$, one block is represented by 1D and the other by the ER random with $q=0$, (b) One block is represented by 1D and the other by ER random with $q>0$, (c) Both the blocks are random, (d) Spacing distribution of layer $2$ when both the blocks are random and $q$ is increased, and (e) Spacing distribution of a multiplex network for different values of  $D_x$. Network size in layer $1$ is $N_1=1000$ and in layer $2$ both blocks are of size $n_1=n_2=500$ such that the size of the network is $N_2=1000$. Average degree in both the layers is $\langle k \rangle=20$.}
\label{fig.10}
\end{figure}


Fig.~\ref{fig.10}(a) presents the spacing distribution of multiplex networks which follows a distribution intermediate between Poisson and GOE statistics when layer $2$ has one block represented by an ER random network. It shows that the presence of a random network even as a subnetwork in one of the layers is enough for an increase in the Brody parameter. When connections are introduced between the blocks with a probability $q=0.002$ which corresponds to on average one connection $(\langle l \rangle=n\times q)$, NNSD of the multiplex network follows GOE statistics (Fig.~\ref{fig.10}(b)). Next, we consider both the blocks in layer $2$ represented by the ER random networks with $p=0.04$ and for $q=0$, the spacing distribution again depicts the GOE statistics (Fig.~\ref{fig.10}(c)). Whereas the spacing distribution of the isolated networks forming the layer $2$ is a superposition of two GOEs for $q=0$ and GOE statistics as connections are introduced between the two blocks $(q>0)$ (Fig.~\ref{fig.10}(d)). For $q=0$ which corresponds to the situation that in the layer $2$, there exists no connections between the blocks. The eigenvalues of the network is union of the eigenvalues of both the blocks due to which NNSD results in superposition of $2$ GOEs \cite{block}. However, when a network consisting of disconnected blocks in the layer $2$ is multiplexed with another network in layer $1$, the adjacency matrix of the entire multiplex network has block matrices along diagonal with off-diagonal entries (inter-layer connections) resulting in GOE statistics of the spectra of the entire multiplex network.
This additional experiment supports the results that presence of a random network even as a subnetwork in one of the layers is enough to make the entire multiplex network sufficiently random to introduce short range correlations in its spectra. On increasing $q$, the NNSD of the multiplex network remains the same and it keeps following GOE statistics while NNSD of the layer $2$ undergoes a transition to the GOE as depicted  by Fig.~\ref{fig.10}(d).
Next, we bring our focus to impact of the arrangement of the inter-layer connections and their strengths on the spectra. We vary $D_x$ from $0.05$ to $1$ and fit the NNSD of the spectra of the entire network using the Brody formula. For very small values of $D_x$, the spacing distribution follows the Poisson statistics and as $D_x$ is increased, a transition from Poisson to GOE is observed (Fig.~\ref{fig.10}(e)).

\begin{figure}[!htb]
    \centering
    
    { \includegraphics[width=0.8 \linewidth]{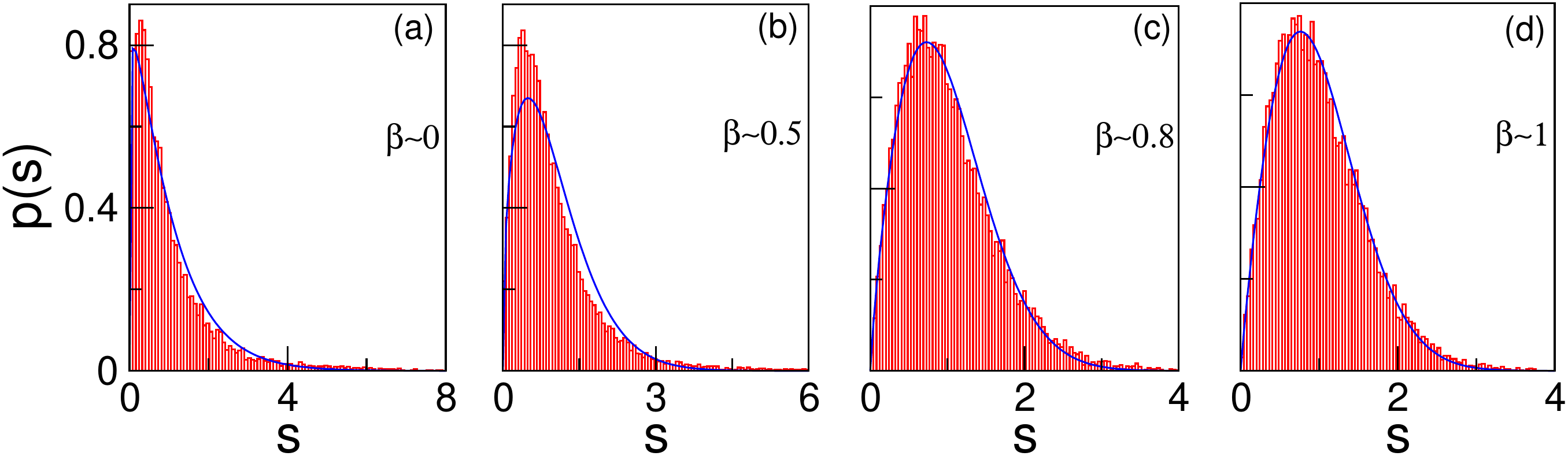}} 
    
     \vspace{0.3cm}
    
    { \includegraphics[width=0.4 \linewidth]{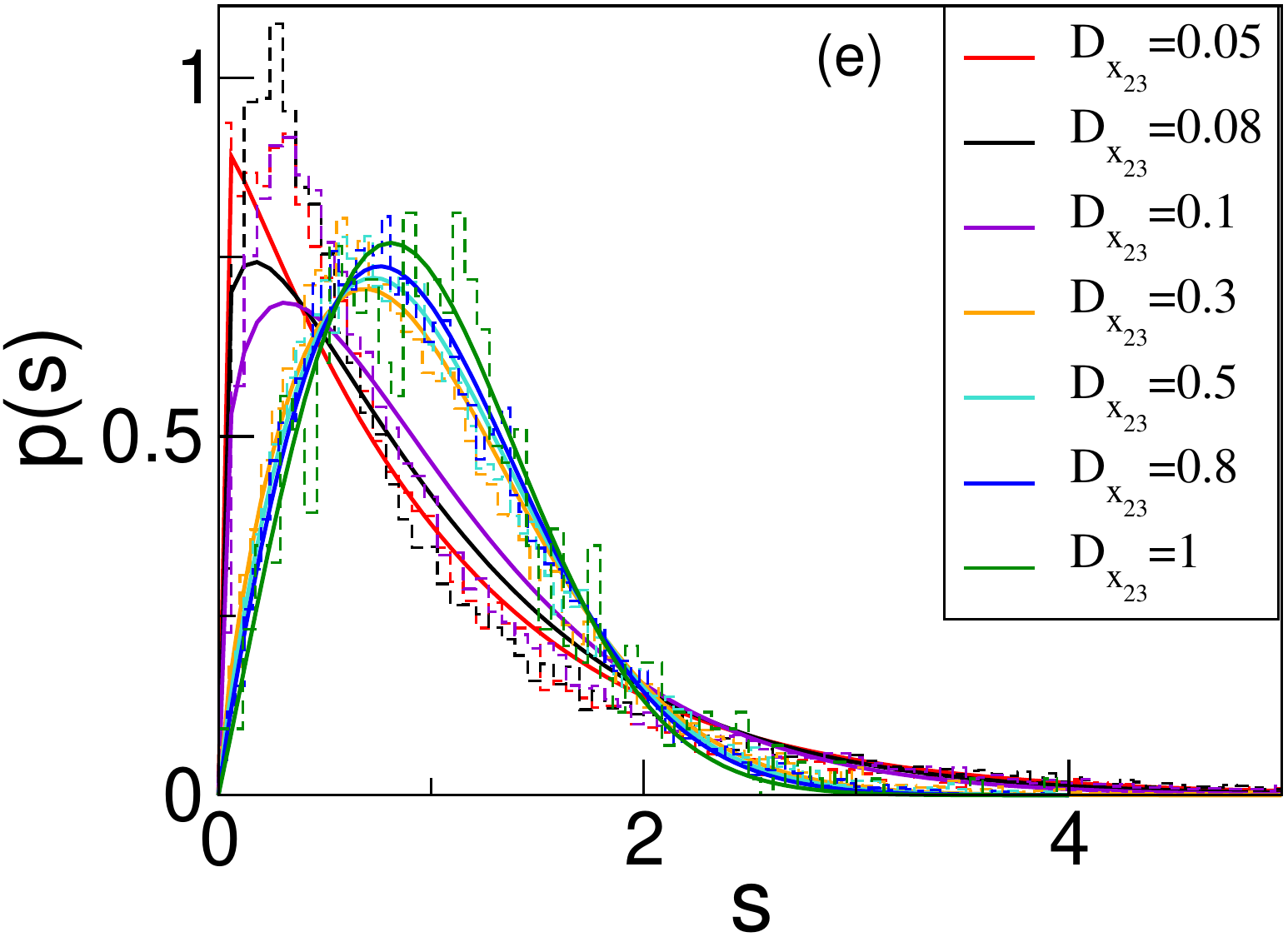}}
 
    \caption{(Color online) Spacing distribution of the multiplex networks. Layer $1$ and $2$ are fixed as 1D lattice, and layer $3$ is formed by the network at different values of the rewiring probability $p_r = 0.0001$ (a), $0.001$ (b), $0.01$ (c), $1$ (d) from left to right. (e) $D_{x_{23}}$ is increased from $0.05$ to $1$.}

\label{fig.11}
\end{figure}  


\subsection{Three layer multiplex network}
Here, we extend our analysis to three layer multiplex networks to analyze the impact of {\it randomness} in only one layer on the spectral properties of the entire multiplex network. The multiplex network for such an experiment is designed as follows: two layers are kept as 1D lattice and through Watts-Strogatz algorithm, the third layer is rewired by changing the value of $p_r$ from $0.0001$ to $1$. The network size and the average degree for all the three layers are taken the same to discard any change arising due to mismatch in the average degree.
First we discuss the effect of randomness in one layer on the entire spectra. It has been shown that random rewiring of the edges, even with a small probability, can lead to drastic changes in the spectral properties from that of the 1D lattice as well as bi-layer multiplex network constituting one such layer. Next, in three-layer multiplex networks, keeping 1D lattice architecture in two of the layers (say layer $1$ and $2$) fixed and rewiring the edges in the third layer, we observed that for even a very small value of the rewiring probability, {\it randomness} of the entire multiplex network is increased reflected by an increase in the value of the Brody parameter (Fig.~\ref{fig.11}(a)-(d)). Initially for the small values of $p_r$, the spacing distribution of the combined multiplex network follows the Poisson statistics. However, as $p_r$ increases, the value  of $\beta$ increases to $1$. It again reflects that a very small random rewiring in only one layer can lead to diminishing of the regularity of the entire multiplex network leading to enhancement in the Brody parameter as also observed for bi-layer multiplex networks.
Next, we study the role of multiplexing strength on the spacing distribution. The three layers multiplex network provides two additional number of parameters, ($D_{x_{12}}$: multiplexing strength between layers $1$ and $2$, $D_{x_{23}}$: multiplexing strength between layers $2$ and $3$). First we vary $D_{x_{23}}$  which is the multiplexing strength between the layers represented by 1D lattice (layer $2$) and the rewired layer ($p_r=1$, layer $3$) while keeping $D_{x_{12}}$, the multiplexing strength between the layers $1$ and $2$ (both representing the 1D lattice) fixed. Fig.~\ref{fig.11}(e) illustrates that as $D_{x_{23}}$ increases, $\beta$ value manifests a consistent increase eventually leading to a smooth transition from the Poisson to the GOE statistics, again similar to the behaviour observed for bi-layer multiplex networks.

\subsection{Impact of edge deletion}

\begin{figure}[t]
\setlength{\abovecaptionskip}{6pt}
\setlength{\belowcaptionskip}{0pt}
\centering
    \includegraphics[width=0.5\textwidth]{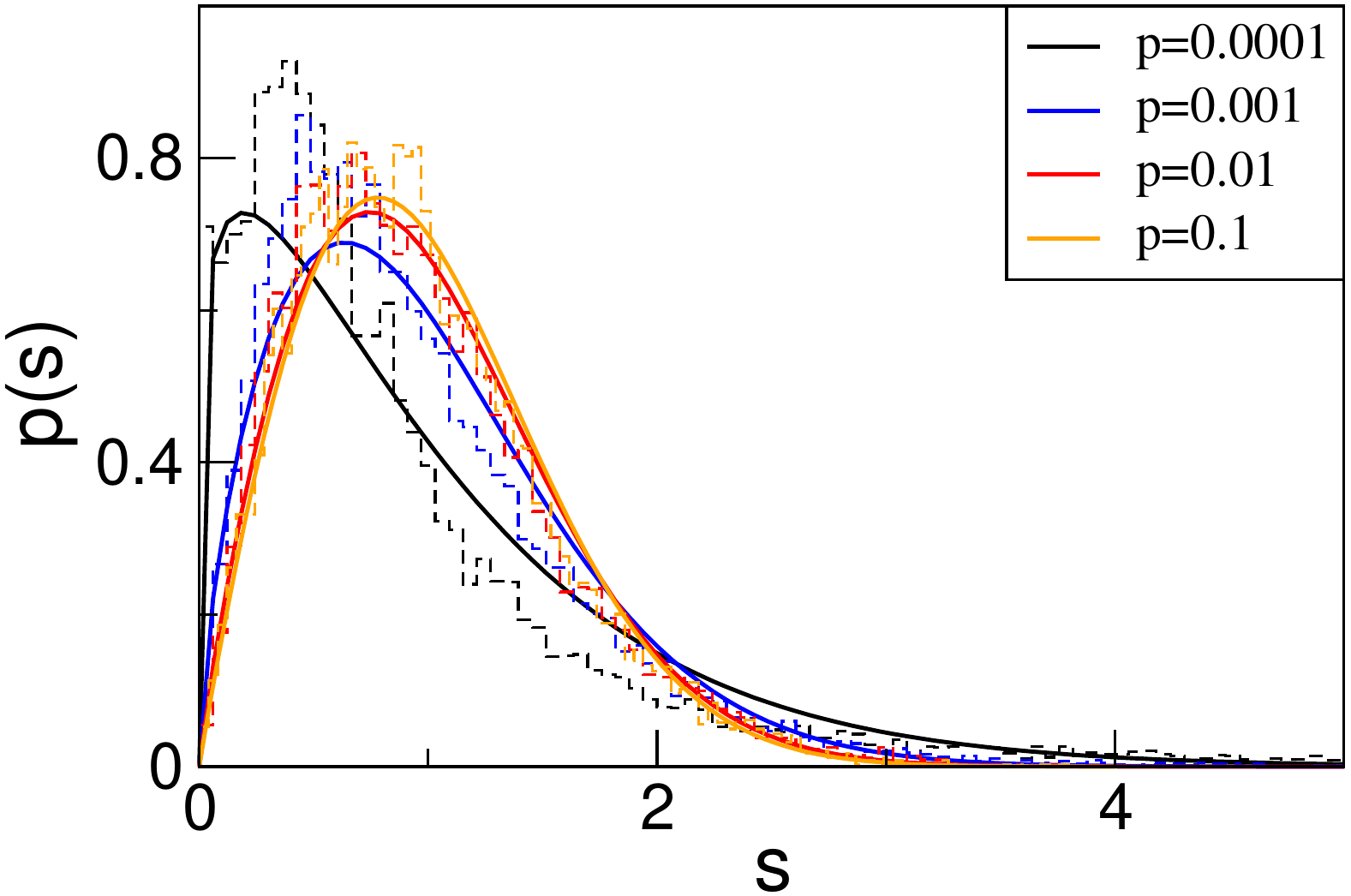}
    \caption{(Color online) Spacing distribution of the multiplex network indicating transition from Poisson to GOE statistics as $p_d$ is increased from $0.0001$ to $0.1$.}
\label{fig.12}
\end{figure}

This section presents results for the impact of edge deletion in one layer on the spectra of entire multiplex networks. Edge deletion in complex networks has been widely investigated area, particularly to study its effect on structural controllability \cite{ed_del1} or to quantify the vulnerability of a complex network under random damages \cite{ed_del2}. We randomly delete edges with a probability $p_d$ in one of the layers (say layer $2$). Note that by doing so, we always ensure that the network in individual layer remains connected. For a fixed value of $p_d$, on average $(N_2*p_d*\langle k_2 \rangle/2)$ edges will be deleted. We present the results for bi-layer multiplex networks in which both layers are represented by 1D lattice. The network parameters are taken the same as of the previous analysis. Fig.~\ref{fig.12} depicts that as the value of $p_d$ increases i.e., as more number of edges are deleted in one of the layers, the spacing distribution of that layer as well as of the entire multiplex network exhibit changes such that $\beta$ changes from $0$ to $1$. For 1D lattice structure, a deletion of the edges leads to distortion of the band structure around the diagonal causing randomness. As anticipated from our observation reported in the previous section, inducing randomness by deleting edges in only one layer is good enough to percolate randomness in the multiplex network reflected  by the spectral properties.
Further, on removing the edges from one of the layers represented by the random network, by keeping the entire  network connected, the spectral statistics of that layer and entire multiplex network remains the same. For the case of a random network, $0$ and $1$ entries are placed randomly in the corresponding adjacency matrix and therefore a random deletion of few edges causes few random entries to become $0$. However, since the average degree of the network reduces as $p_d$ increases, if average degree of a network reduces to $\langle k \rangle \leq 1$, spacing distribution of the corresponding network follows poisson statistics \cite{RMT_random}. Also, network becomes disconnected, whereas we have restricted our analyses to the connected networks.
\begin{figure}[t]
\setlength{\abovecaptionskip}{6pt}
\setlength{\belowcaptionskip}{0pt}
\centering
    \includegraphics[width=0.8\textwidth]{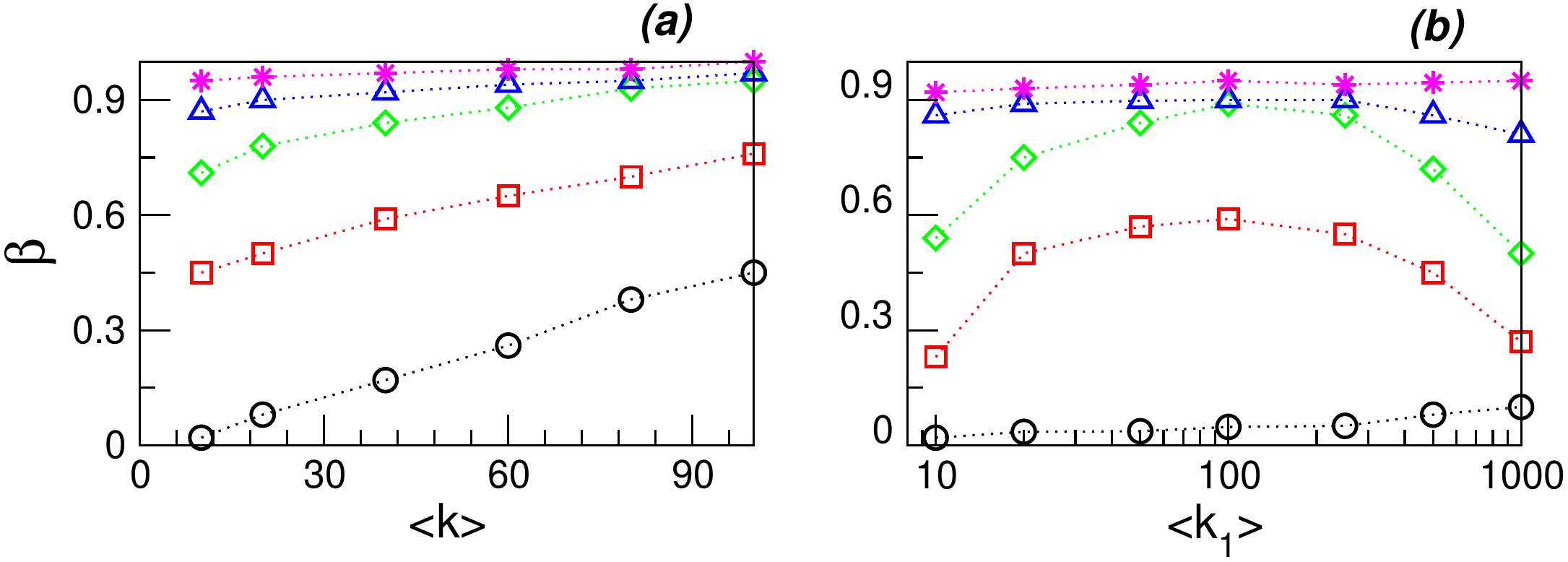}
    \caption{Brody parameter $\beta$ for different values of the rewiring probability $p_{r}$ for layer 2 (while layer 1 is fixed to the 1D lattice). (a) average degree $\langle k \rangle=\langle k_1 \rangle=\langle k_2 \rangle$ is varied in both the layers, (b) as average degree of 1D lattice representing layer 1 ($\langle k_1 \rangle$) is increased (logarithmic scale). All results shown here are for bi-layer networks with $N_{1}=N_{2}=1000$ averaged over $20$ random realisations of the networks. The $\color{black} \circ$, $\color{red} \square$, $\color{green} \diamond$, $\color{blue} \triangle$ and $\color{pink} \star$ correspond to $p_r=0.0001, 0.001, 0.01, 0.1$ and $1$.}
\label{fig.13}
\end{figure}

\begin{figure}[t]
\setlength{\abovecaptionskip}{6pt}
\setlength{\belowcaptionskip}{0pt}
\centering
    \includegraphics[width=0.7\textwidth]{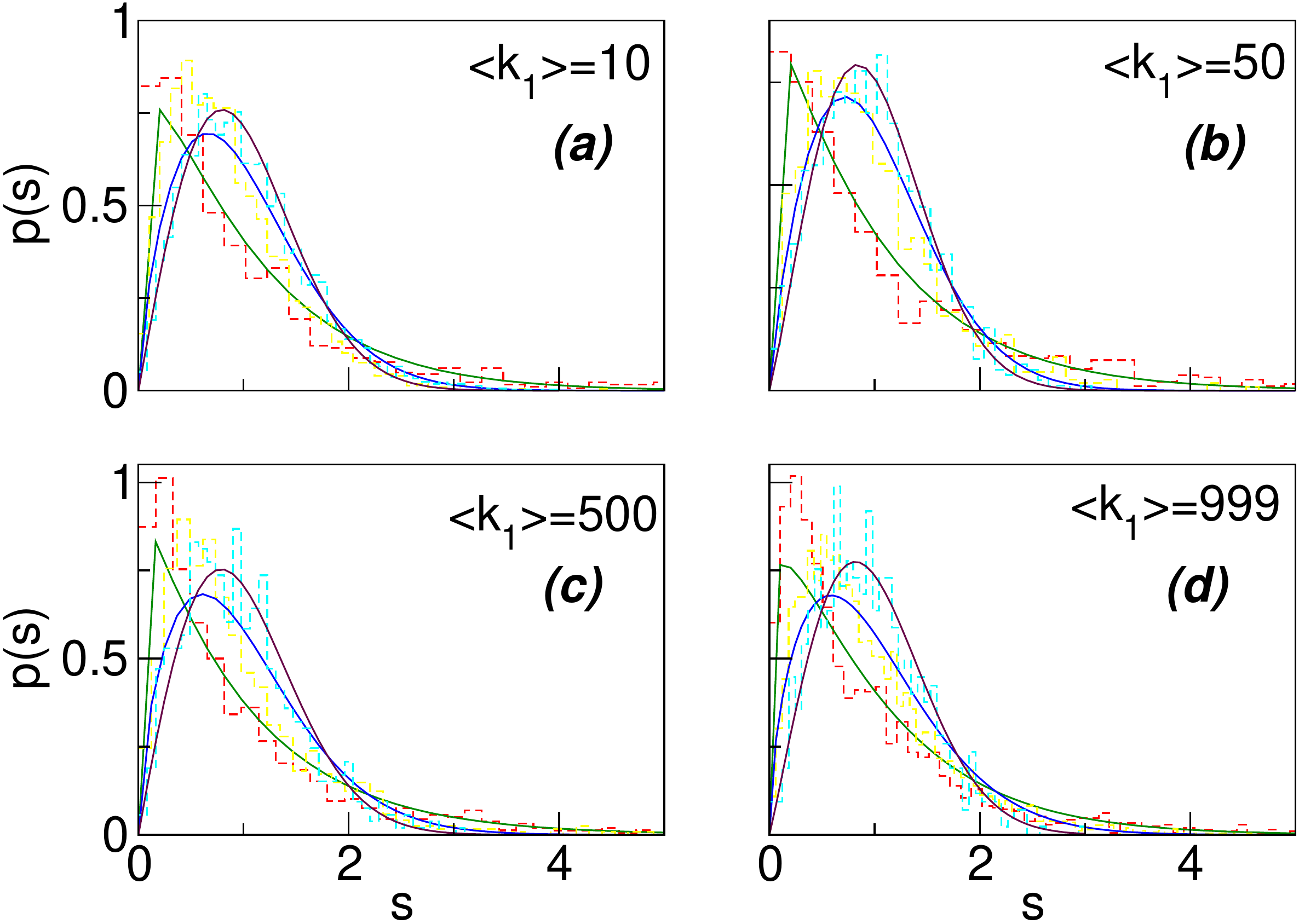}
    \caption{(Color online) Brody parameter $\beta$ as a function of the rewiring probabilities $p_{r}$ for different $\langle k_1 \rangle$. All the results are for bi-layer networks with size $N_{1}=N_{2}=1000$ and $\langle k_{2}\rangle=20$ averaged over 20 random realisations. The first layer is represented by 1D lattice while the second layer undergoes rewiring at different $p_r$ represented by $ \color{green} -$, $ \color{blue} -$ and $ \color{Fuchsia} -$.}
\label{fig.14}
\end{figure}

\subsection{Impact of increase in the average degree}
Next, we increase the average degree of both the layers. Fig.~\ref{fig.13}(a) illustrates variation in the Brody parameter $\beta$ with a change in the average degree of the layers. The multiplex network here has regular 1D lattice forming its first layer and SW network with different values of the rewiring probability as the second layer. As the average degree increases, the Brody parameter $\beta$ increases for a fixed rewiring probability value (Fig.~\ref{fig.13}(a)), which is not surprising as on increasing $\langle k \rangle$ by keeping $N$ and $p_{r}$ fixed, the average number of randomly rewired edges given by ($N \times p_{r} \times k/2$) increases leading to the increase in the {\it randomness} of the entire network. 

\subsubsection{Degree mismatch between the layers}
Fig.~\ref{fig.13}(b) shows the impact of an increase in the average degree of 1D lattice forming the layer $1$ on the spectra of the multiplex networks when the average degree of the layer $2$ is kept fixed. We consider various different values of the rewiring probability for the second layer. Increasing the average degree of the 1D lattice network forming the layer $1$ is also equivalent to an increase in the regularity of this layer as well as of that of the entire multiplex network for a fixed $p_r$ value (of the second layer). Fig.~\ref{fig.13}(b) plots the change in the Brody parameter of the multiplex network as $\langle k_{1} \rangle$ increases from $10$ corresponding to the sparse networks to $N-1$ corresponding to a globally connected network. It can be seen that for $p_{r}=0.001, 0.01, 0.1$ as $\langle k_{1} \rangle$ increases, $\beta$ increases for  $\langle k_{1}\rangle \le 100$ after which $\beta$ starts decreasing. This is not the case for $p_{r}=0.0001$ as such a small rewiring probability does not affect the spectral statistics of the entire multiplex network, and therefore an increase in the regularity also does not affect the spectra of the network. Also, at $p_{r}=1$ for which the network forming the layer $2$ becomes completely random, no effect of the increase in the regularity is observed as the value of $\beta$ remains at the value $1$. These experiments confirm that one of the layers having the random network architecture, increasing the regularity by increasing the average degree of the 1D lattice layer does not affect the short range correlation in the eigenvalues of the entire multiplex network. Also, for various different values of the average degree of the layer represented by the 1D lattice, with an increase in $p_r$ values, NNSD of the entire multiplex network shows a smooth transition from the Poisson to the GOE statistics (Fig.~\ref{fig.14}). This also suggests that in a multiplex network consisting of a globally connected network forming its one of the layers, rewiring of the edges of the other layer can also lead to changes in the spectral properties of the entire multiplex network.


\section{Conclusions}
The present study has numerically investigated the eigenvalues and the NNSD for multiplex networks. We first investigated the impact of structural randomness (in one of its layers only) introduced by the random rewiring of the edges using SW model on the spectra of the entire multiplex networks under the RMT framework. Upon multiplexing two networks having the same architecture does not yield the spectral and NNSD distribution different than those of the individual layer. The spacing distribution follows the Poisson statistics for both the layers being regular and GOE statistics for both the layers being random, which is expected. For the multiplex networks consisting of different network architecture representing its different layers, depending upon structural properties of the individual layer, the spectra of the entire multiplex network may have properties different than either one or both of its layers. The spectral density of the multiplex networks consisting of one layer being regular and one being random follows the semicircular distribution accompanied with many sharp peaks. Whereas for this setup, the spacing distribution follows the GOE statistics, an index of randomness in complex networks, suggesting that there exist short range correlations in the eigenvalues which are arising here due to the {\it randomness} in the adjacency matrix corresponding to one layer only. Further, upon one layer represented by the ER random network, irrespective of the structural properties of the other layer, the structural randomness arising due to the first layer completely dominates the spacing distribution of the entire multiplex networks, and NNSD follows GOE statistics. In fact, if one of the layers of a multiplex network has structural randomness of SW networks, the NNSD keeps following GOE statistics irrespective of the network architecture of another layer, demonstrating that such small randomness which is enough to make a 1D lattice to a SW network is enough to introduce short range correlations in the eigenvalues of the entire multiplex networks. Importantly, these impacts of structural changes in one layer on the spectral properties of the entire multiplex networks can be controlled by tuning of the multiplexing strength $D_x$. For the small multiplexing strengths, $D_x\le0.1$, there is continuous rise in the randomness of the multiplex networks in terms of short range correlations in the eigenvalues captured through the Brody parameter, which increases as the inter-layer connection strengthens. For $D_x\geq0.1$, since NNSD follows GOE statistics already, there is no impact of a further increase in the multiplexing strength on NNSD. Also, using extended analysis of multiplex networks consisting of three layers, we have shown that {\it randomness} in only one layer can impart drastic effect on the spectral fluctuations as observed for bi-layer multiplex network. Different ways of inducing randomness as well as different types of random networks are considered in one of the layers for consistency of results.

The results presented here reflect that the RMT analysis of multiplex networks is capable of capturing the impact of {\it randomness} in one layer on the spectra of the entire underlying network consisting of that layer. The framework is useful to understand dynamical response of a system against such perturbation, which make connectivity between the edges more random. 
Further, this work has only focused on the short range correlations in the eigenvalues. It would be interesting to apply other techniques of RMT such as long range correlations among the eigenvalues (spectral rigidity) and eigenvector localization to achieve further insight.
Furthermore, since real-world networks can show very different spectral features than those of the corresponding model networks, the analysis performed here builds up a platform to investigate the spectra of real-world multilayer networks under the RMT framework.


\section*{Acknowledgments}

SJ and Tanu acknowledge Govt. of India, BRNS Grant No.37(3)/14/11/2018-BRNS/37131 for financial support and JRF fellowship, respectively. We thank Shashi Chandralal Srivastava (VECC) for interesting discussions throughout the project, and Ranveer Singh (IIT Indore) for useful suggestions.



\begin{thebibliography}{99}
\bibitem{CLR} In case of both the layers represented by 1D lattice, to avoid the degeneracy of eigenvalues \cite{Dup_nodes1} of a multiplex network, the 1D lattice  in second layer is rewired at very small probability $p_{r}=0.0001$. An isolated 1D lattice rewired at $p_{r}=0.0001$ also follows poisson distribution.
\end{thebibliography}
\end{document}